\documentclass[twocolumn]{bytedance_seed}

\usepackage[toc,page,header]{appendix}

\usepackage{minitoc}

\usepackage[english]{babel}
\usepackage{blindtext}
\usepackage{caption}
\usepackage{color}
\usepackage{multirow}
\usepackage{url}
\usepackage{xurl}
\usepackage{rotating}
\usepackage{array}
\usepackage{algorithmicx,algorithm}
\usepackage[noend]{algpseudocode}
\usepackage{enumitem}
\usepackage{xspace}
\usepackage{latexsym}
\usepackage{amsmath}
\usepackage{booktabs}
\usepackage{tabularx}
\usepackage{balance}
\usepackage{graphicx}
\usepackage{svg}
\usepackage{pifont}
\svgsetup{
    inkscapepath=i/svg-inkscape/
}
\svgpath{{svg/}}
\usepackage{etoolbox}
\usepackage{colortbl}

\newcommand{\parabf}[1]{\medskip\noindent\textbf{#1}}

\newcommand{\paraf}[1]{\noindent\textbf{#1}}
\newcommand{\cut}[1]{}

\newcommand{\sysname}{MegaScale-Infer\xspace}
\newcommand{\topk}{top-$k$\xspace}

\newcommand{\scalemoe}{Scaled-MoE\xspace}

\title{MegaScale-Infer: Serving Mixture-of-Experts at Scale with Disaggregated Expert Parallelism}

\author[1,2,\circ,*]{Ruidong Zhu}
\author[1,\circ]{Ziheng Jiang}
\author[1,2,\circ,*]{Chao Jin}
\author[1]{Peng Wu}
\author[1]{Cesar A. Stuardo}
\author[1]{Dongyang Wang}
\author[1]{Xinlei Zhang}
\author[1]{Huaping Zhou}
\author[1]{Haoran Wei}
\author[1]{Yang Cheng}
\author[1]{Jianzhe Xiao}
\author[1]{Xinyi Zhang}
\author[1]{Lingjun Liu}
\author[1]{Haibin Lin}
\author[1]{Li-Wen Chang}
\author[1]{Jianxi Ye}
\author[1]{Xiao Yu}
\author[2, \dagger]{Xuanzhe Liu}
\author[2, \dagger]{Xin Jin}
\author[1, \dagger]{Xin Liu}

\affiliation[1]{ByteDance Seed}
\affiliation[2]{Peking University}
\contribution[\circ]{Equal Contribution}

\contribution[*]{Work done at ByteDance Seed}
\contribution[\dagger]{Corresponding authors}

\abstract{
    Mixture-of-Experts (MoE) showcases tremendous potential to scale large language models (LLMs) with enhanced performance and reduced computational complexity. However, its sparsely activated architecture shifts feed-forward networks (FFNs) from being compute-intensive to memory-intensive during inference, leading to substantially lower GPU utilization and increased operational costs.

    We present \sysname, an efficient and cost-effective system for serving large-scale MoE models. \sysname disaggregates attention and FFN modules within each model layer, enabling independent scaling, tailored parallelism strategies, and heterogeneous deployment for both modules.
    To fully exploit disaggregation in the presence of MoE's sparsity, \sysname introduces \textit{ping-pong pipeline parallelism}, which partitions a request batch into micro-batches and shuttles them between attention and FFNs for inference. Combined with distinct model parallelism for each module, \sysname effectively hides communication overhead and maximizes GPU utilization. 
    To adapt to disaggregated attention and FFN modules and minimize data transmission overhead (e.g., token dispatch), \sysname provides a high-performance M2N communication library that eliminates unnecessary GPU-to-CPU data copies, group initialization overhead, and GPU synchronization. Experimental results indicate that \sysname achieves up to 1.90$\times$ higher per-GPU throughput than state-of-the-art solutions.
}

\correspondence{Xuanzhe Liu, Xin Jin, Xin Liu}

\begin{document}
\maketitle


\section{Introduction}
\label{sec:introduction}

\begin{figure*}[t!]
    \centering
    \includegraphics[width=0.9\linewidth]{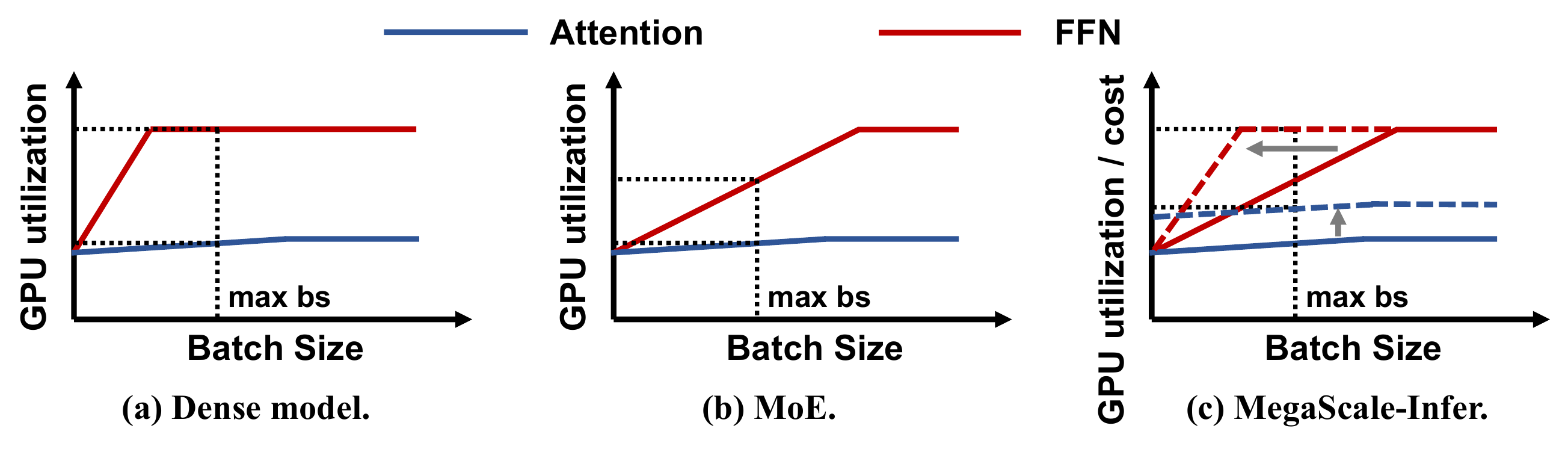}
    \vspace{-0.1in}
    \caption{GPU utilization of Attention and FFN vs. batch size in dense model, MoE, and \sysname during decoding.}
    \vspace{-0.1in}
    \label{fig:intro:gpu_util}
\end{figure*}

Large language models (LLMs), such as GPT-4~\cite{achiam2023gpt}, Claude~\cite{claude3}, and Llama~\cite{touvron2023llama, touvron2023llama2, dubey2024llama}, have revolutionized the field of artificial intelligence, demonstrating remarkable proficiency in numerous domains. These models have not only enhanced existing technologies like search engines~\cite{chatgpt-search} but have also paved the way for innovative applications in areas like universal chatbots~\cite{chatgpt, gemini} and programming assistants~\cite{github-copilot, cursor}. 

As the effectiveness of LLMs increasingly depends on the escalation of model parameters, there is a growing imperative to scale up these models~\cite{jiang2024megascale, liu2024deepseek}. 
Due to the sparse activation architecture, mixture-of-experts (MoE) models~\cite{lepikhin2020gshard, rajbhandari2022deepspeed} are a practical choice for scaling. MoE dynamically routes input tokens to a subset of feed-forward networks (FFNs), which are known as experts, rather than engaging all FFNs (i.e., all parameters). This design enables sub-linear scaling of required FLOPs as the number of experts and model size increases, significantly reducing computational complexity without compromising model quality.

Unfortunately, reduced computational complexity does not necessarily translate into lower computational costs in practical serving scenarios. This discrepancy arises from the mismatch between the characteristics of LLM inference and the compute capabilities of GPUs, a problem that becomes increasingly pronounced with growing MoE sparsity. Figure~\ref{fig:intro:gpu_util} demonstrates this issue. Specifically, an LLM consists of multiple layers of attention and FFN modules. During the decoding phase, which dominates the LLM inference process~\cite{kwon2023efficient}, the GPU utilization of attention modules remains low because they must access the intermediate states (i.e., key-value cache) of all previous tokens. Conversely, FFN modules achieve high GPU utilization as the number of tokens increases.

However, GPU memory limitations and response latency constraints impose an upper bound on the number of tokens that can be processed simultaneously (i.e., batch size). For dense models, which contain one FFN module per layer, this maximum batch size allows the FFN to fully utilize the GPUs' compute capabilities. 
In MoE models, however, larger model sizes are often accompanied by more experts and higher sparsity, meaning that fewer tokens---less than a quarter, or even an order of magnitude less---are assigned to each expert within the same batch size. 
As depicted in Figure~\ref{fig:intro:gpu_util}(b), the increased sparsity lowers the GPU utilization of FFN modules, rendering them no longer compute-intensive, and resulting in unnecessary computational costs.

A natural solution is to disaggregate attention from the LLM inference process and replicate attention modules to increase the decoding batch size for FFN modules. This approach is adopted by Infinite-LLM~\cite{lin2024infinite}, which focuses on optimizing dense model inference in long-context scenarios. In such cases, GPU memory capacity, rather than sparsity, is the primary constraint, and the communication pattern is relatively simple compared to the \topk selection in MoE. Consequently, its solution is less effective in addressing the unique challenges of MoE inference.

We present \sysname, an efficient and cost-effective system designed for large-scale MoE serving. \sysname disaggregates the attention and expert modules, 
assigning them to separate GPUs---a strategy we term \textit{disaggregated expert parallelism}. Our approach offers two major benefits. 
First, it enables independent scaling of each module with customized model parallelism strategies. Specifically, attention modules are replicated using data parallelism, while FFN modules are scaled with expert parallelism. By consolidating requests from multiple attention replicas, the GPU utilization of each expert increases significantly as the batch size per attention replica grows. Second, it enables the deployment of attention and FFN modules on heterogeneous GPUs to fully leverage their different capabilities and achieve lower costs. For example, attention modules can be deployed on GPUs with more cost-effective memory capacity and bandwidth, while FFN modules can utilize GPUs with more affordable compute capability. As shown in Figure~\ref{fig:intro:gpu_util}(c), FFN can easily become compute-intensive in \sysname, while attention achieves higher GPU utilization per unit cost under heterogeneous deployment.

Disaggregated expert parallelism introduces two new technical challenges. First, the disaggregation architecture causes the attention and FFN modules to be idle for a batch when the other is computing or when they are waiting for tokens. We design a ping-pong pipeline parallelism strategy that splits a batch of requests into multiple micro-batches to keep the attention and FFN busy and hide the communication overhead. Furthermore, the effectiveness of the ping-pong pipeline parallelism strategy depends on certain conditions, such as similar computation time for attention and FFN. To fill the pipeline and maintain high GPU utilization, \sysname optimizes the model parallelism strategy for each module based on a performance model specifically designed for disaggregated MoE serving. 

Second, the arbitrary parallelism configuration of the attention and FFN modules transforms the original All2All communication between them for token routing into M2N communication, where M and N represent the number of senders and receivers, respectively. Based on our observations about the performance shortcomings of popular communication libraries~\cite{nccl} in the context of this specific communication pattern, we develop a high-performance M2N communication library with a focus on reducing operational overhead and improving communication stability.

We implement \sysname and evaluate it using MoE models with sizes ranging from 132 to 317 billion parameters. The experimental results show that \sysname outperforms state-of-the-art LLM serving systems by up to 1.9$\times$ in per-GPU decoding throughput.
We also conduct experiments on a heterogeneous cluster, where \sysname achieves 1.7$\times$ higher throughput per unit cost. Compared to NCCL~\cite{nccl}, a widely-used communication library, \sysname's M2N communication achieves 4.2$\times$ higher throughput and 68.2\% lower latency.
\sysname has already been deployed in the company's inference services and reduces the serving cost by 1.5--2.0$\times$.

In summary, we make the following contributions.
\begin{itemize}[leftmargin=*]
    \item We present \sysname, a system for efficiently serving large-scale MoE-based LLMs. Leveraging insights into the characteristics of Transformer and MoE, we employ a disaggregated approach for the attention and FFN modules. This approach offers dual advantages: it enables tailored parallelism strategies and independent hardware selection, thereby optimizing system efficiency and cost-effectiveness.
    
    \item In order to support the disaggregated serving architecture at scale, we present a ping-pong pipeline parallelism strategy to utilize GPU compute capabilities and hide communication, and develop a high-performance M2N communication library to enhance network performance.
    
    \item Our experiments demonstrate significant improvements in throughput and cost-effectiveness with our system's unique capabilities. \sysname achieves up to 1.90$\times$ and 1.86$\times$ per-cost decoding throughput against state-of-the-art LLM serving systems on homogeneous and heterogeneous clusters, respectively.
\end{itemize}
This work does not raise any ethical issues.
\section{Background and Motivation}
\label{sec:background}

\subsection{LLM Inference Characteristics}
\label{sec:background:inference}

A Transformer-based LLM typically consists of multiple layers, with each layer containing an attention module and an FFN module.
Unlike traditional DNN inference, LLM inference follows an autoregressive pattern.
It takes a sequence of input tokens, known as a prompt, as input and goes through the attention and FFN modules for multiple iterations to generate output tokens.
In the prefill phase or the first iteration, the model computes the attention between each pair of tokens in the prompt to produce the first output token.
During this iteration, intermediate representations, or key-value (KV) cache, are stored for each token.
These cached representations are then used in the subsequent iterations to calculate the attention.
In the following decoding iterations, the LLM generates the next token by computing the attention between the newly generated token and all previous tokens.

The autoregressive generation pattern makes the attention module compute-intensive during the prefill phase and memory-intensive during the decoding phase.
Even with request batching~\cite{yu2022orca, triton-server}, a widely-used optimization in efficient LLM serving, attention during the decoding phase remains the same memory access intensity.
This is because each request has its own KV cache of input and previously generated tokens, which is different from each other.
In the decoding iteration, each request must access its respective KV cache.
In contrast, the computation of FFN only requires loading the corresponding model weights from GPU memory to SRAM, which can be shared across all tokens from different requests.
Consequently, as presented in Figure~\ref{fig:intro:gpu_util}(a), batching is only efficient for FFNs to reuse model parameters and improve GPU utilization.

\subsection{LLM Serving at Scale}
\label{sec:background:scale}

The scaling law~\cite{kaplan2020scaling} highlights the significance of model size as a key determinant of the model capability. To achieve state-of-the-art model capability, many efforts~\cite{jiang2024megascale, liu2024deepseek} have been invested in scaling LLMs to hundreds of billions of parameters. Due to the large model size, serving these models necessitates both algorithmic and system optimizations.

\parabf{Mixture of experts.} From an algorithmic perspective, mixture-of-experts (MoE) models show significant potential in enhancing the performance of LLMs with sub-linear scaling computational complexity and are gaining popularity in large-scale model implementations~\cite{lepikhin2020gshard, fedus2022switch, deepseekai2024deepseekv2, liu2024deepseek}. We focus on MoE in Transformer-based LLMs in this work.

\begin{figure}[t]
    \centering
    \includegraphics[width=\linewidth]{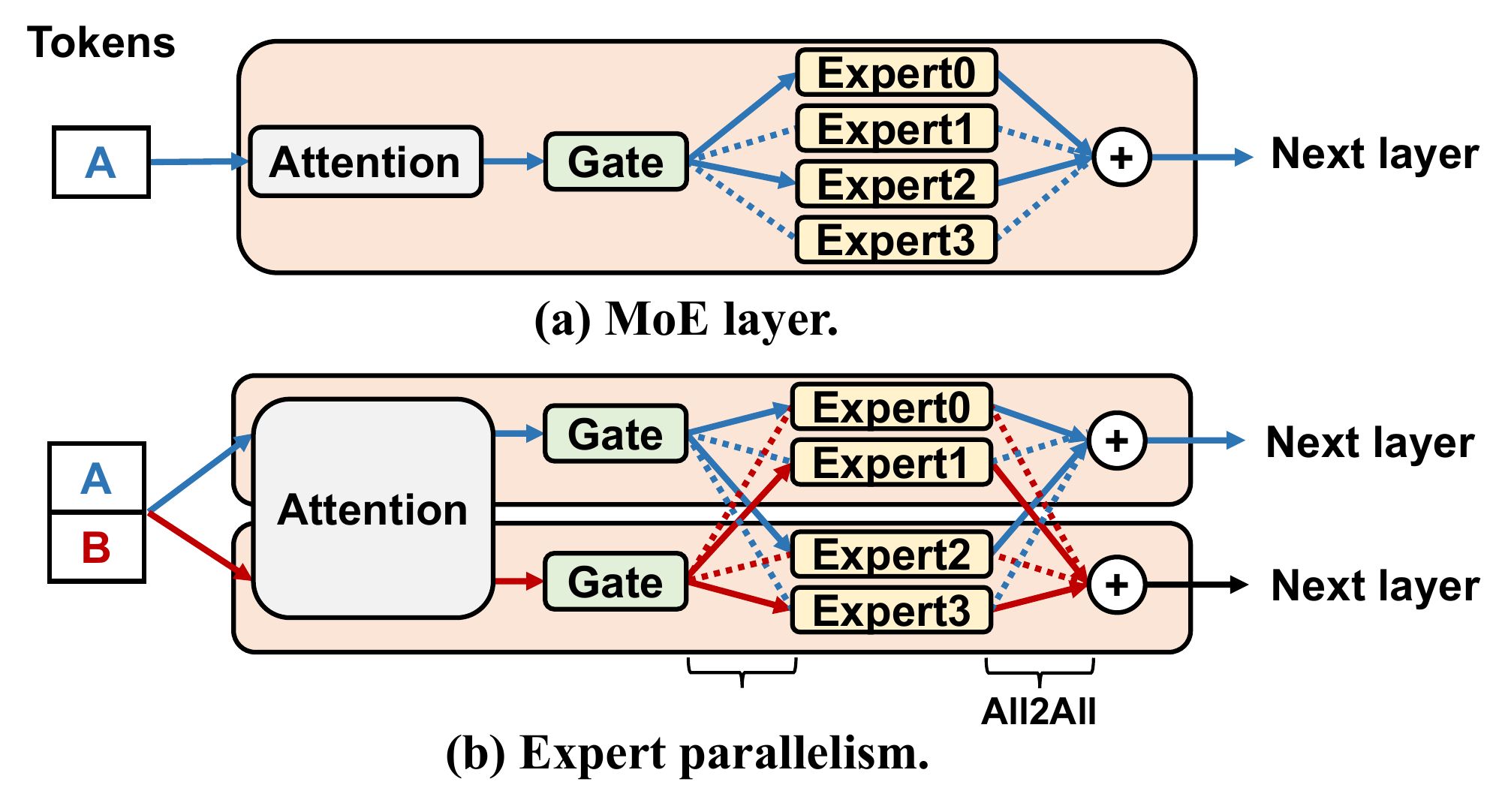}
    \vspace{-0.2in}
    \caption{MoE and expert parallelism.}
    \vspace{-0.1in}
    \label{fig:background:moe}
\end{figure}

MoE models replace the feed-forward network (FFN) layer with an MoE layer, which consists of multiple FFNs acting as experts, as shown in Figure~\ref{fig:background:moe}(a). A gating network within the MoE layer routes input tokens to a subset of these experts, i.e., top-k experts, based on matrix multiplication between each token's embedding vector and the gating network's trainable parameters. The final output of the MoE layer is a weighted sum of the selected experts' outputs. The sparse nature of MoE allows for scaling the model size by increasing the number of experts without linearly raising computational costs.
For instance, Mixtral 8x22B~\cite{mixtral822} has around 141B parameters, but its active parameters for each token are only approximately 39B with top-2 expert selection.

\parabf{Model parallelism.} From a systems perspective, serving large-scale LLMs requires a distributed approach due to the limited memory and compute capacity of a single device. Model parallelism distributes model parameters across multiple devices to improve efficiency. Tensor parallelism~\cite{shoeybi2019megatron} (TP) partitions compute-intensive operators like matrix multiplications to accelerate computation, but it introduces substantial communication overhead. Thus, tensor parallelism is usually confined to a single node with multiple GPUs, where intra-node NVLink bandwidth is typically much higher than inter-node network bandwidth. Pipeline parallelism~\cite{huang2019gpipe} divides model layers into stages, each running on a device to form a pipeline. This method slightly increases inference time due to inter-stage communication but scales serving throughput linearly with each additional stage.

A parallelism strategy specialized for MoE named expert parallelism (EP) is also widely applied in MoE serving~\cite{rajbhandari2022deepspeed}. As shown in Figure~\ref{fig:background:moe}(b), each device only contains some of the experts in expert parallelism. Consequently, the forward pass of an MoE layer requires two all-to-all communications: one to send input tokens to the experts selected by the gating network, and the other to send the processed tokens back. In EP, the computation of each expert involves complete matrix multiplication, which is more conducive to GPU computation compared to TP, where a single matrix multiplication is split across multiple GPUs.
The potential issue of EP is load imbalance between experts and the increased communication volume as the number of top-k experts grows.
Therefore, whether TP or EP benefits FFN more depends highly on the structure of MoE models and the real-time workload.

\subsection{Problems in Large-scale MoE Serving}
As demonstrated in \S\ref{sec:background:inference}, the memory-intensive attention operation during the decoding phase leads to low GPU utilization, while FFNs can achieve high efficiency through request batching. However, the sparsity of MoE alters this situation.
Although the sparsity enables sub-linear scaling of computational complexity, it significantly decreases the inference efficiency.
Figure~\ref{fig:intro:gpu_util}(b) presents a schematic diagram of the impact.
Given a request batch during the decoding phase, each expert processes only a portion of them, resulting in a smaller batch size for FFNs, thereby lowering the GPU utilization.

Take Mixtral 8x22B as a more concrete example.
Assume that we use NVIDIA A100-SXM-80GB GPUs, which have a computational power of 312 TFLOPS and memory bandwidth of 2 TB/s, to serve this model with the bfloat16 datatype.
The floating point operations required for a $b\times h$ to $h \times n$ GEMM (General Matrix to Matrix Multiplication) are $2bhn$, where $b$ and $h$ represent the decoding batch size and the model's hidden dimension size, respectively.
The number of parameters this GEMM needs to access is $hn$, and the data volume is $2hn$ for bfloat16.
Let the GPU's floating point compute capability be $F$ and the memory bandwidth be $B$.
According to the roofline model~\cite{williams2009roofline}, a GPU requires that $\frac{2bhn}{F} \geq \frac{2hn}{B}$, i.e., $b \geq \frac{F}{B}$, to fully utilize its matrix multiplication capability.
For an A100 GPU, the batch size at least needs to be 156 tokens ($\frac{312 TFLOPS}{2TB/s}$).
However, given a batch size of 156, the average number of decoding tokens dispatched to each expert is $156 \times topk / \#expert = 156 \times 2 / 8 = 39$, with the theoretical Model Flops Utilization (MFU) for FFN modules of $2 / 8 =25\%$.
Formally, the theoretical relationship between batch size and FFN's GPU utilization for dense models is $util = min(\frac{B}{F} b, 1)$, but for MoE, it is $util=min(\frac{topk}{\#expert} \frac{B}{F} b, 1)$.

Ideally, we can enhance the inference efficiency by increasing the batch size, but in practice, there are many factors that constrain the batch size.
For instance, a larger batch size may compromise the requirement of low latency in online model serving.
Additionally, the GPU memory constraint for KV cache limits the batch size growth.
Especially for large-scale MoE, the GPU memory becomes more scarce, resulting in a smaller maximum batch size.
Although enlarging the model parallelism with more GPUs may allow a larger batch size, it also introduces more communication overhead.

\begin{figure}[t]
    \centering
    \includegraphics[width=\linewidth]{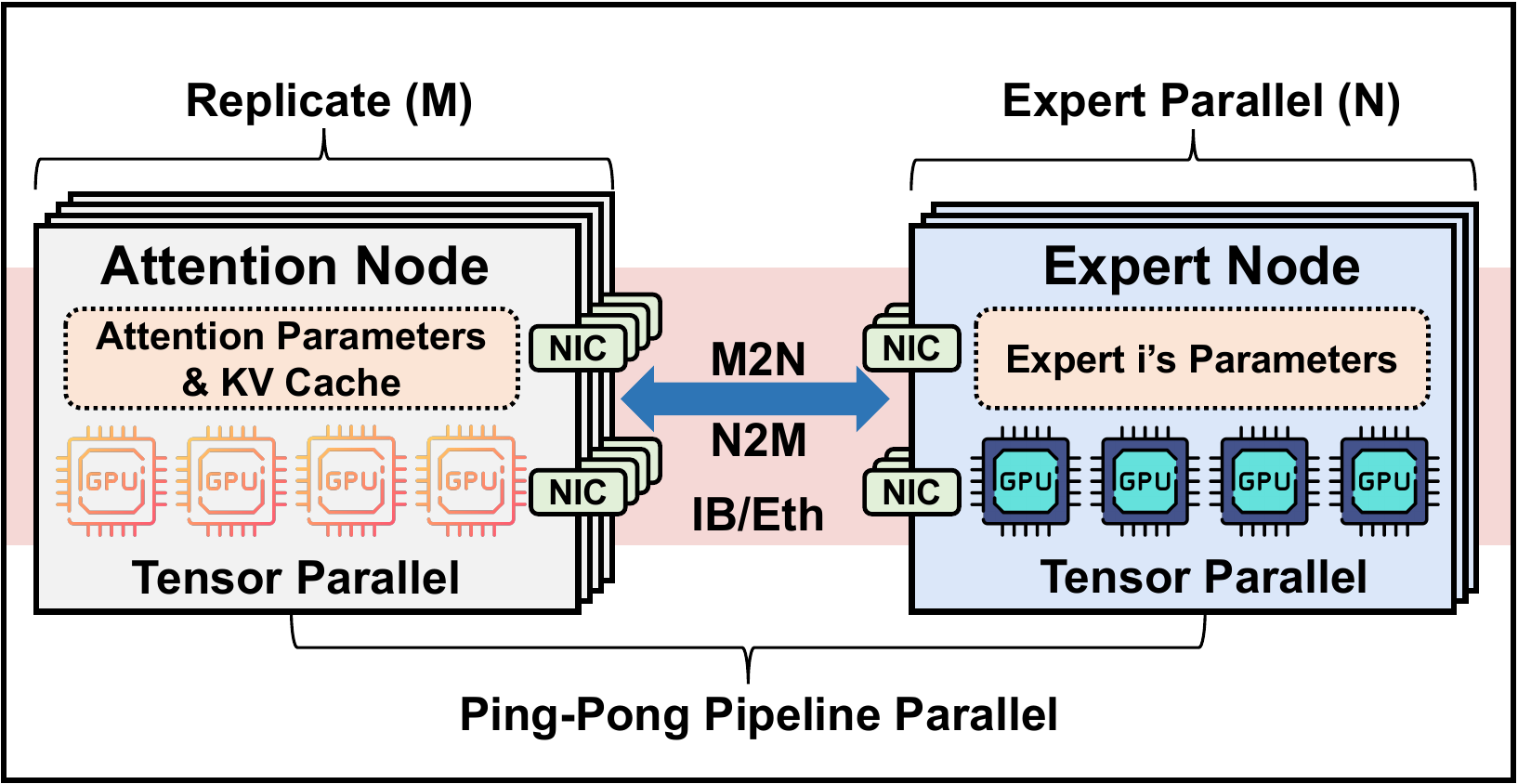}
    \vspace{-0.1in}
    \caption{\sysname runtime instance architecture.}
    \vspace{-0.1in}
    \label{fig:overview}
\end{figure}

\subsection{Opportunities and Challenges}
\label{sec:background:opportunity}

To address the inefficiency caused by MoE sparsity, we find that disaggregating the attention modules and FFN modules naturally provides two key advantages: 
\begin{itemize}[leftmargin=*]
    \item \textbf{Independent scaling.} This allows us to scale serving instances with attention modules independently, aggregating decoding requests for each FFN module. This makes the FFN module compute-intensive and achieves optimal GPU utilization.
    \item \textbf{Heterogeneous deployment.} The disaggregated architecture naturally separates the deployment for attention and FFN modules, allowing for the use of the most cost-effective GPUs for each. It also opens up opportunities to use specialized hardware and software to separately accelerate attention and FFN computation.
\end{itemize}
There are two main technical challenges to realize efficient disaggregation of attention and FFN. 
First, since each token must repeatedly and sequentially pass through the attention and FFN modules, disaggregating these two components introduces idle periods. Specifically, the attention modules remain idle while the FFN modules are performing computations, and vice versa. Both modules can also experience idle time while waiting for outputs to be transmitted over the network. Therefore, a ping-pong pipeline must be established between the attention and FFN modules to ensure continuous utilization. Furthermore, this pipeline should be meticulously co-designed with the model parallelism strategies of each module to maximize GPU utilization while adhering to latency requirements.

Second, the independent scaling enabled by disaggregation requires M2N and N2M communication between M attention GPUs and N expert GPUs, replacing the traditional All-to-All communication used in each MoE layer. However, directly leveraging peer-to-peer communication primitives from existing libraries results in significant performance degradation, highlighting the need for a specialized communication library tailored to the M2N pattern.

\section{\sysname Overview}
\label{sec:overview}

In this work, we present \sysname, a system designed for efficiently serving MoE-based LLM at scale.
Following prior work~\cite{zhong2024distserve, patel2023splitwise}, \sysname decouples prefill and decoding into separate clusters to eliminate their interference and meet their respective latency requirements.
In this paper, we focus on the decoding phase, aiming to address its inefficiency.
Figure~\ref{fig:overview} illustrates the overall architecture of a \sysname runtime instance serving a single model replica during the decoding phase.
By disaggregating the attention and FFN modules onto separate attention and expert nodes, respectively, \sysname allows for independent scaling and heterogeneous deployment of attention and FFN, significantly enhancing system efficiency and reducing serving costs.

\begin{table}[t!]
    \centering
    \resizebox{0.9\linewidth}{!} {
    \begin{tabular}{ll}
        \toprule
        \textbf{Symbol} & \textbf{Description} \\
        \midrule
        $B$ & Global batch size per instance\\
        $m$ & \#micro-batches\\
        $b_a, b_e$ & Micro-batch size per node\\
        $h$ & Hidden size of the LLM\\
        $h'$ & Intermediate dimension size of FFN \\
        $g$ & Number of query heads per group in GQA \\
        $L$ & \#layers of the LLM\\
        $s$ & Average sequence length in a batch\\
        $K$ & number of selected experts for each token\\
        $E$ & \#experts / \#expert nodes per instance\\
        $n_a$ & \#attention nodes per instance\\
        $tp_a, tp_e$ & TP size for attention and expert nodes\\
        $N_m$ & \#micro-batches limit per instance\\
        $M_a, M_e$ & \#GPUs per node limit for attention and expert\\
        $C_a, C_e$ & GPU memory capacity for attention and expert\\
        $P_a, P_e$ & Parameter size of attention and one expert\\
        $T_a, T_e$ & Computation time of one micro-batch\\
        $T_c$ & Communication time of one micro-batch\\
        $tpuc$ & throughput per unit cost\\
        \bottomrule
    \end{tabular}
    }
    \caption{Key notations.}
    \vspace{-0.15in}
    \label{tab:design:notations}
\end{table}

\parabf{Disaggregated expert parallelism.} To facilitate large-scale MoE serving, \sysname employs a hybrid parallelism strategy called \emph{disaggregated expert parallelism}. Each expert node typically consists of 1-8 GPUs within a single physical server and stores the parameters of one expert. All expert nodes together form an expert parallelism group. The parameters of the attention module (e.g., weight matrices for QKV and output projection) are replicated on each attention node, where the key-value caches are also stored. 
Tensor parallelism is employed within each attention/expert node to leverage high-bandwidth connectivity between GPUs (e.g., NVLink). 
\sysname also designs a ping-pong pipeline parallelism strategy tailored to the disaggregated architecture, feeding micro-batches of requests into attention and expert nodes to keep them busy during communication or while awaiting results from other nodes.
\sysname determines the detailed deployment plan based on a performance model designed for the disaggregated expert parallelism.

\parabf{High-performance M2N communication.} \sysname employs a customized M2N communication library to transfer the intermediate outputs between each pair of attention nodes and expert nodes.
To achieve efficient and stable data transmission, the library removes unnecessary GPU-to-CPU data copies, group initialization overhead, and GPU synchronization.
It also proposes traffic-oriented optimizations specific to this scenario.
\section{Disaggregated Expert Parallelism}
\label{sec:ep}

In this section, we present the design of ping-pong pipeline parallelism and the approach to generating the deployment plan of \sysname.
Given the MoE model, workload characteristics (e.g., sequence lengths), available hardware, and latency requirements, \sysname determines the deployment plan by specifying $(i)$ the respective parallelism strategies for attention and experts, $(ii)$ the number of micro-batches for the ping-pong pipeline, $(iii)$ the maximum batch size, and $(iv)$ the hardware setup for deployment. Our goal is to identify the deployment plan that maximizes throughput per unit cost (e.g., dollar). Table~\ref{tab:design:notations} lists the key notations in our discussion.
We assume the model uses grouped-query attention (GQA)~\cite{ainslie2023gqa}, which is the most popular method for attention.

\subsection{Ping-Pong Pipeline Parallelism}
\label{sec:ep:pipeline}

\begin{figure}[t]
    \centering
    \includegraphics[width=\linewidth]{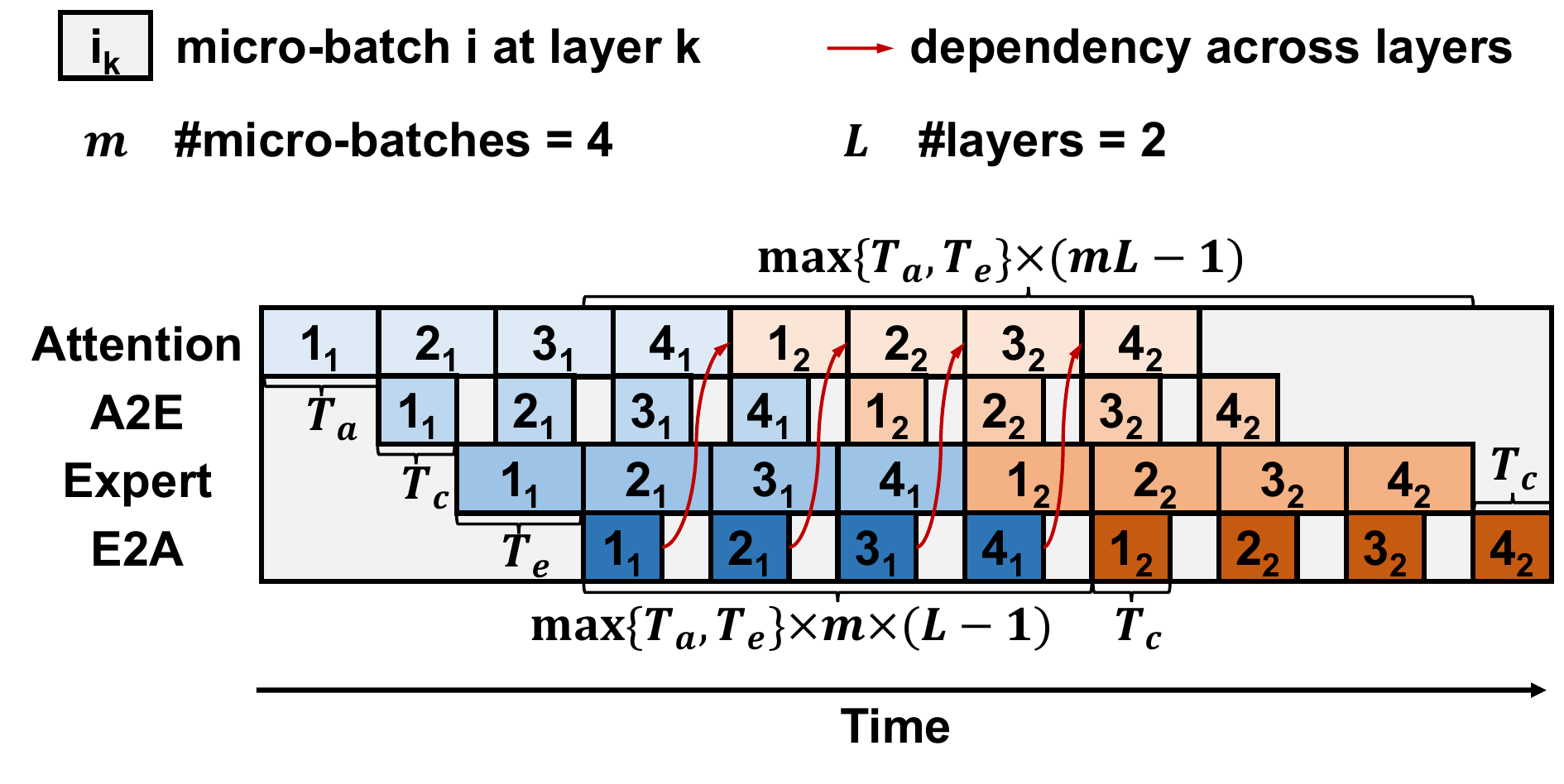}
    \vspace{-0.2in}
    \caption{Illustration of ping-pong pipeline parallelism.}
    \vspace{-0.1in}
    \label{fig:ep:pingpong_pp}
\end{figure}

As we decouple the FFN modules from the attention modules, using a single batch of requests would result in idle time for both the attention nodes and the expert nodes when the other module is busy. GPUs also remain idle during the inter-node communication. To address this problem, as illustrated in Figure~\ref{fig:ep:pingpong_pp}, we split a batch of requests into $m$ micro-batches, creating a ping-pong pipeline between the attention nodes and expert nodes. 
These nodes perform the forward pass of the micro-batches and exchange intermediate results twice in each MoE layer. This setup allows the forward computation to cover the communication overhead, thereby achieving higher GPU utilization.

Let $T_a$ and $T_e$ represent the computation time of a micro-batch on an attention node and an expert node, respectively. We define $T_f = \max\{T_a, T_e\}$ as the maximum of these two values. $T_c$ denotes both the communication time from attention nodes to expert nodes and vice versa, as the two bi-directional communications share the same network configuration. Our objective is to overlap communication with computation, keeping the GPUs fully utilized. The necessary conditions to achieve this are 
\begin{align}
    T_a & \approx T_e, \label{formula:balance} \\
    T_c & < T_f, \label{formula:hide1} \\
    m \times T_f & \geq 2\times(T_f + T_c). \label{formula:hide2}
\end{align}

Constraint~\ref{formula:balance} aims to minimize the GPU idle time caused by computation dependencies across MoE layers. Constraint~\ref{formula:hide1} and constraint~\ref{formula:hide2} describe methods for hiding communication overhead. Specifically, the communication time for a single micro-batch must be shorter than the forward computation time of attention and experts, and the forward time of one MoE layer for the global batch on each node must be sufficient to cover the time required for a single micro-batch to pass through the layer. We can then obtain the minimum number of micro-batches needed using formula $m \geq 2\times (1 + \frac{T_c}{T_f})$, where $0 < \frac{T_c}{T_f} < 1$. For deployments with fast communication ($T_c < \frac{1}{2} T_f$), at least 3 micro-batches are required. For those with relatively slower communication, at least 4 micro-batches are required.

Let the number of MoE layers be $L$. As illustrated in Figure~\ref{fig:ep:pingpong_pp}, considering the imbalanced computation between attention nodes and expert nodes, the decoding iteration latency of one micro-batch can be estimated as
\begin{align}
    (T_a+T_e+2T_c)+mT_f(L-1) \leq T_{iter} \leq mT_fL.
\end{align}
The total iteration latency of the global batch is 
\begin{align}
    T_{total} = (T_a+T_e+2T_c)+T_f(mL-1).
\end{align}

\begin{algorithm}[t!]
    \caption{Deployment Plan Search for Decoding Phase}
    \label{alg:disaggr:plan}
    \raggedright
    \textbf{Input:} MoE model $G$, $C_a$, $C_e$, $N_m$, $M_a$, $M_e$\\
    \textbf{Output:} the optimal deployment plan $plan^*$
    \begin{algorithmic}[1]
        \State $plan^* \gets \emptyset$
        \For{$tp_e \in \{1, 2, \ldots, M_e \}$}
            \For{$tp_a \in \{1, 2, \ldots, M_a\}$}
                \If{$tp_a \times C_a > P_a$ and $tp_e \times C_e > P_e$}
                    \State $n_a \gets $ \Call{balance}{$G$, $tp_a$, $tp_e$}
                    \For{$m \in \{3, 4, \ldots, N_m\}$}
                        \State $plan \gets \{(tp_e, E), (tp_a, n_a), m\}$
                        \State $B$, $tpuc \gets $ \Call{simulate}{$G, plan, SLO$}
                        \State $plan \gets plan \cup \{B, tpuc\}$
                        \If{$plan^*.tpuc < plan.tpuc$}
                            \State $plan^* \gets plan$
                        \EndIf
                    \EndFor
                \EndIf
            \EndFor
        \EndFor
    \end{algorithmic}
\end{algorithm}

\subsection{Deployment Plan Search}
\label{sec:ep:formulation}

Considering ping-pong pipeline parallelism, the search space of \sysname deployment plan includes the tensor parallelism sizes for attention nodes ($tp_a$) and expert nodes ($tp_e$), the number of attention nodes ($n_a$), the number of micro-batches, and the global batch size ($B$). Our objective is to minimize the throughput per unit cost while adhering to the SLO constraint. Algorithm~\ref{alg:disaggr:plan} shows the pseudo-code for searching the optimal deployment plan given hardware setup and model configurations. It enumerates the feasible $tp_a$ and $tp_e$, subject to GPU memory capacity limit. For each pair of $tp_a$ and $tp_e$, it calculates the number of attention nodes to balance the computation time as closely as possible according to constraint~\ref{formula:balance}. The algorithm then compares the throughput per unit cost among deployment plans with varying numbers of micro-batches. Using the SIMULATE function, it determines the maximum global batch size that meets the SLO through binary search and obtains the optimal plan.

The complexity of Algorithm~\ref{alg:disaggr:plan} is $O(M^2N_m)$, with $M$ as the GPU limit per server and $N_m$ as the maximum number of micro-batches. Typically, $M$ has four choices (e.g., $\{1, 2, 4, 8\}$) in modern clusters. We set $N_m$ to four because splitting into too many micro-batches reduces GEMM efficiency in expert nodes and thus increases the latency. Consequently, the search space remains manageable.

\begin{table}[t!]
    \centering
    \arrayrulewidth=0.5pt
    \extrarowheight=1pt
    \resizebox{0.85\linewidth}{!} {
        \begin{tabular}{c|c|c}
            \arrayrulecolor{black}\hline
            \arrayrulecolor{black}\hline
            GEMM Name & Shape of Input & Shape of Param.\\
            \hline
            QKV Project & $(b_a, h)$ & $(h, h(1+2/g)/tp_a)$ \\
            Attn Output & $(b_a, h/tp_a)$ & $(h/tp_a, h)$ \\
            FFN Input & $(b_e, h)$ & $(h, h'/tp_e)$ \\
            FFN Output & $(b_e, h'/tp_e)$ & $(h'/tp_e, h)$ \\
            \arrayrulecolor{black}\hline
            \arrayrulecolor{black}\hline
        \end{tabular}
    }
    \caption{GEMMs used in MoE inference.}
    \vspace{-0.15in}
    \label{tab:design:gemm}
\end{table}

\parabf{Performance simulation.} We then dive into the MoE layers to analyze the simulation of $T_a$, $T_e$, and $T_c$. $T_a$ includes two GEMMs: QKV Project and Attn Output, while $T_e$ includes another two GEMMs: FFN Input and FFN Output. Their input and parameter shapes are shown in Table~\ref{tab:design:gemm}. The arithmetic intensity of attention GEMMs and FFN GEMMs are $O(b_a)$ and $O(b_e)$, respectively, with the relationship $b_a \times m \times n_a = b_e \times m \times E / K = B$. The attention module is memory-intensive since it needs to access the KV cache of all tokens in the batch. Let the average sequence length be $s$, the KV cache access time is nearly proportional to $b_as$. The tensor parallelism synchronization time is $O(b_ah(tp_a - 1)/tp_a)$. Thus, we can model $T_a$ as $k_1b_a+k_2$ and model $T_e$ as $k_3b_e+k_4$ similarly, where $k_i$ values can be obtained through profiling and interpolation as prior work does~\cite{zhong2024distserve}. Consequently, $n_a = (b_eE)/(b_aK)$ can be set as $(k_1E)/(k_3K)$ to balance $T_a$ and $T_e$.

As for $T_c$, it equals the maximum time between sending and receiving. We profile the relationship between network bandwidth utilization and message size to estimate $T_c$. Specifically, 
\begin{align}
    T_c = \max\{ \frac{b_ahK/tp_a}{W_a\times Util(b_ahK/tp_a)}, \frac{b_eh/tp_e}{W_e\times Util(b_eh/tp_e)} \},
\end{align}
where $W_a$ and $W_e$ represent the network bandwidth per GPU on attention and expert nodes, respectively.

In addition to constraint~\ref{formula:balance}, \ref{formula:hide1}, and~\ref{formula:hide2}, there are two constraints in the search process:
\begin{align}
    T_{iter} & \leq SLO,\\
    4mb_ashL/g + 2P_a & < tp_aC_a. \label{formula:kv_cache}
\end{align}
Constraint~\ref{formula:kv_cache} represents the GPU memory capacity limit for bfloat16 KV cache size. And the throughput per unit cost is $\frac{B/T_{total}}{tp_an_aCost_a + tp_eECost_e}$.

\begin{table}[t!]
    \centering
    \arrayrulewidth=0.5pt
    \extrarowheight=1pt
    \resizebox{\linewidth}{!} {
        \begin{tabular}{c|c|c|c|c|c|c|c}
            \arrayrulecolor{black}\hline
            \arrayrulecolor{black}\hline
            \multirow{2}{*}{\begin{tabular}[c]{@{}c@{}}Accelerator\end{tabular}} & \multirow{2}{*}{\begin{tabular}[c]{@{}c@{}}Price\end{tabular}} & \multirow{2}{*}{\begin{tabular}[c]{@{}c@{}}Cap.\\(GB)\end{tabular}} & \multirow{2}{*}{\begin{tabular}[c]{@{}c@{}}Bw.\\(GB/s)\end{tabular}} & \multirow{2}{*}{\begin{tabular}[c]{@{}c@{}}Comp.\\(TFLOPS) \end{tabular}} &
            \multicolumn{3}{c}{Performance per Cost} \\ 
            \cline{6-8} 
            &  & & & & GB & GB/s & TFLOPS \\ 
            \hline
            L20 & 1.00 & 48  & 864 & 119.5 & 48 & 864  & 119.5  \\ 
            H800 & 5.28 & 80  & 3430.4 & 989 & 15.2 & 649.7  & 187.3  \\ 
            A800 & 2.26 & 80  & 2039 & 312 & 35.4 & 902.2 & 138.1  \\ 
            H20 & 1.85 & 96 & 4096 & 148 & \textbf{51.9} & \textbf{2214.1} & 80.0 \\
            L40S & 1.08 & 48 & 864 & 362 & 44.4 & 800.0 & \textbf{335.2} \\
            \arrayrulecolor{black}\hline
            \arrayrulecolor{black}\hline
        \end{tabular}
    }
    \caption{Performance specifications and cost-effectiveness of different hardware. Prices are normalized by L20.}
    \vspace{-0.15in}
    \label{tab:design:hetero}
\end{table}

\subsection{Heterogeneous Deployment}
\label{sec:ep:heter}

\sysname supports a heterogeneous hardware setup for attention nodes and expert nodes. Specifically, we use GPUs with higher per-cost memory bandwidth and larger per-cost memory capacity for attention nodes, as these nodes are memory-intensive, spending most of their time on memory access and requiring significant storage for the KV cache. Similarly, for expert nodes, which are compute-intensive, we use GPUs with higher cost-effectiveness in compute capability.

Table~\ref{tab:design:hetero} lists the performance specifications, prices, and corresponding ratios for a selection of NVIDIA GPUs. We enumerate the scenarios of using each type of GPU as the hardware for attention or expert nodes to determine the optimal deployment plan.
Intuitively, H20 is more suitable for attention due to its large memory capacity and high memory bandwidth per unit cost.
Meanwhile, the L40S GPU is more cost-effective for experts.

Heterogeneous deployment can also reduce energy consumption by utilizing hardware with lower energy consumption per unit of compute or bandwidth.
For example, the H20 and L40S GPUs have maximum power consumptions of 500W and 350W, respectively.
Under the same power consumption, the L40S offers higher compute performance, while the H20 provides higher bandwidth.
We demonstrate the improvements in both cost and energy efficiency achieved through heterogeneous deployment in \S\ref{sec:evaluation:e2e}.

\section{High-Performance M2N Communication}
\label{sec:m2n}

In MoE inference, token dispatch and aggregation (i.e., communication between attention and FFN modules) rely on peer-to-peer primitives, as the destinations are determined dynamically. 
To motivate the need for a custom communication library for token dispatch, we start by highlighting the limitations of the existing solution, NCCL~\cite{nccl}. Specifically, we compare NCCL to perftest~\cite{perftest}, a networking micro-benchmark designed to measure latency and throughput from the perspective of a simple CPU client, with convenient support for GPU memory buffers as both sources and destinations. 
We use perftest as a baseline to establish the lower bound of achievable latency, with data dynamically dispatched in a manner similar to token routing in MoE inference.
Figure~\ref{fig:m2n:baseline} presents the observed latency when a single sender transmits 128K bytes to each receiver in N, where |N| = \{8, 16, 32\}. 
Based on this experiment, we derive the following observations:

\begin{figure}[t]
    \centering
    \includegraphics[width=0.9\linewidth]{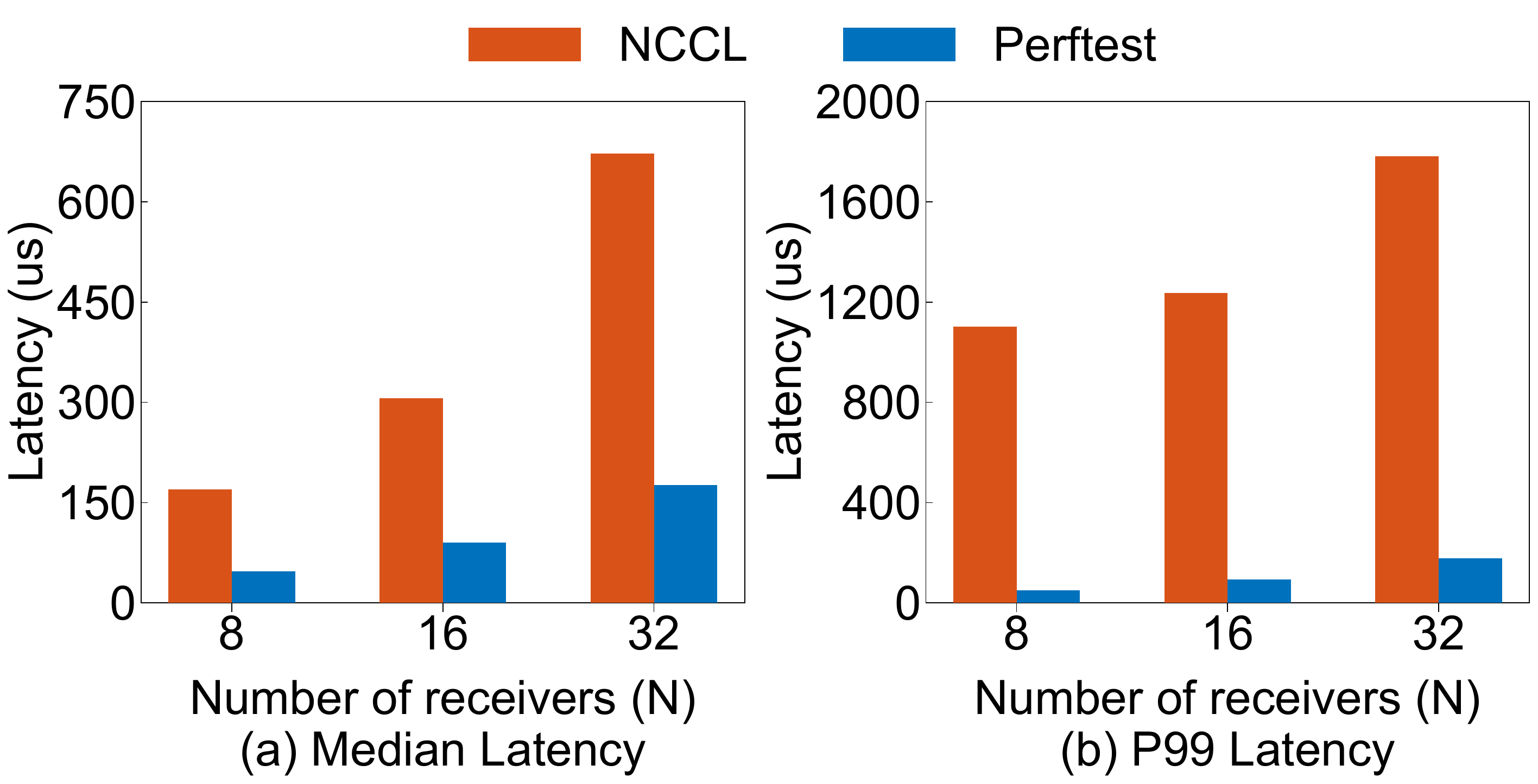}
    \vspace{-0.1in}
    \caption{One-to-N latency: a single sender sends 128K bytes to each receiver in N, where |N| = \{8, 16, 32\}.}
    \vspace{-0.1in}
    \label{fig:m2n:baseline}
\end{figure}

\begin{itemize}[leftmargin=*]
    \item \textbf{High additional overheads.} Figure~\ref{fig:m2n:baseline}(a) shows the median latency for both alternatives. While the scaling trends appear to follow similar patterns, the latency of NCCL significantly exceeds that of the baseline.
    
    \item \textbf{Instability at higher percentiles.} The performance issue highlighted in Figure~\ref{fig:m2n:baseline}(a) consistently exacerbates at higher percentiles, as shown in Figure~\ref{fig:m2n:baseline}(b). At the 99th percentile, the baseline experiences only a slight increase in latency, whereas NCCL exhibits a significant surge, particularly when scaling to 32 or more receivers.
\end{itemize}

The underlying causes of these issues are multifaceted, stemming from specific design choices in NCCL that are not well-suited for this particular use case.
Regarding overhead, we first note that NCCL networking requires intermediate copies \cite{nccl-proxy} to transfer data from the GPU memory to the CPU proxy, which performs the network operations. While features like user buffer registration \cite{nccl-buffer} aim to reduce these copies, they do not fully eliminate them. Second, peer-to-peer group operations \cite{nccl-groups} are processed in batches of at most $8$ operations, which causes a harmful effect as the number of receivers scales. 
Third, as a general-purpose collective communication library, NCCL incurs overhead from general group operation setup, including preparing and launching a batch of N send operations, internal handling and verifications, etc. While these steps are essential for ensuring broad applicability, they introduce unnecessary latency and can be optimized, though not entirely eliminated.
Regarding stability, this issue is much more complex and can arise from multiple sources, including OS, memory, networking, and GPU thermal differences~\cite{sinha2022gpuvariability,xin2024gpuvariability, beckman2006osnoise, hoefler2010osnoise, gunawi2018failslow, tang2018osnoise, zheng2022osnoise}.
Previous studies have highlighted that common sources of instability often arise from GPU synchronization operations and device memory accesses~\cite{zhou2021gpunoise, hao2023gpunoise}, both of which are prevalent in NCCL but absent in the baseline.

\begin{figure}[t]
    \centering
    \includegraphics[height=2.25in]{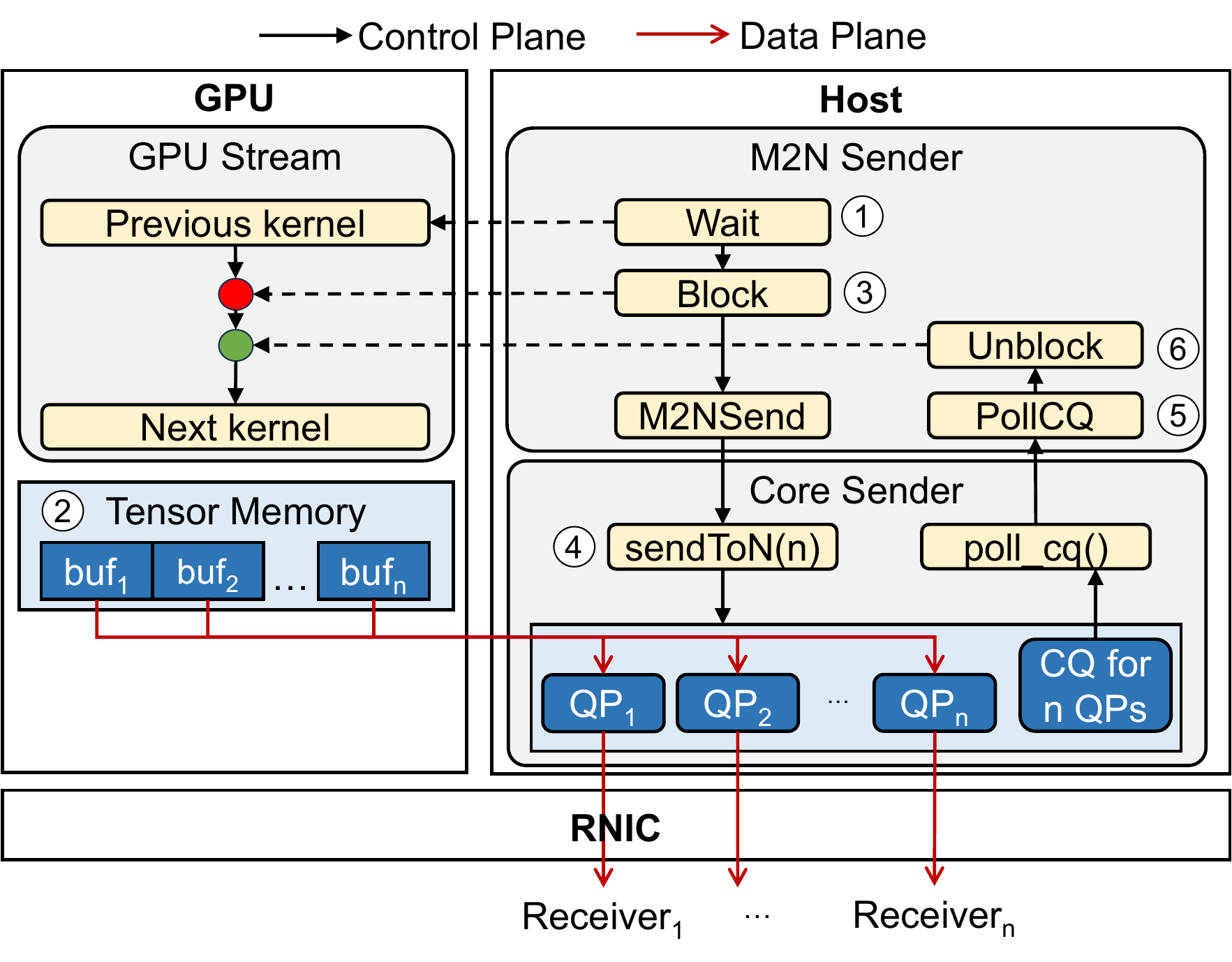}
    \vspace{-0.1in}
    \caption{M2N Sender components and interactions.}
    \vspace{-0.1in}
    \label{fig:m2n:sender}
\end{figure}

\begin{figure}[t]
    \centering
    \includegraphics[height=2.25in]{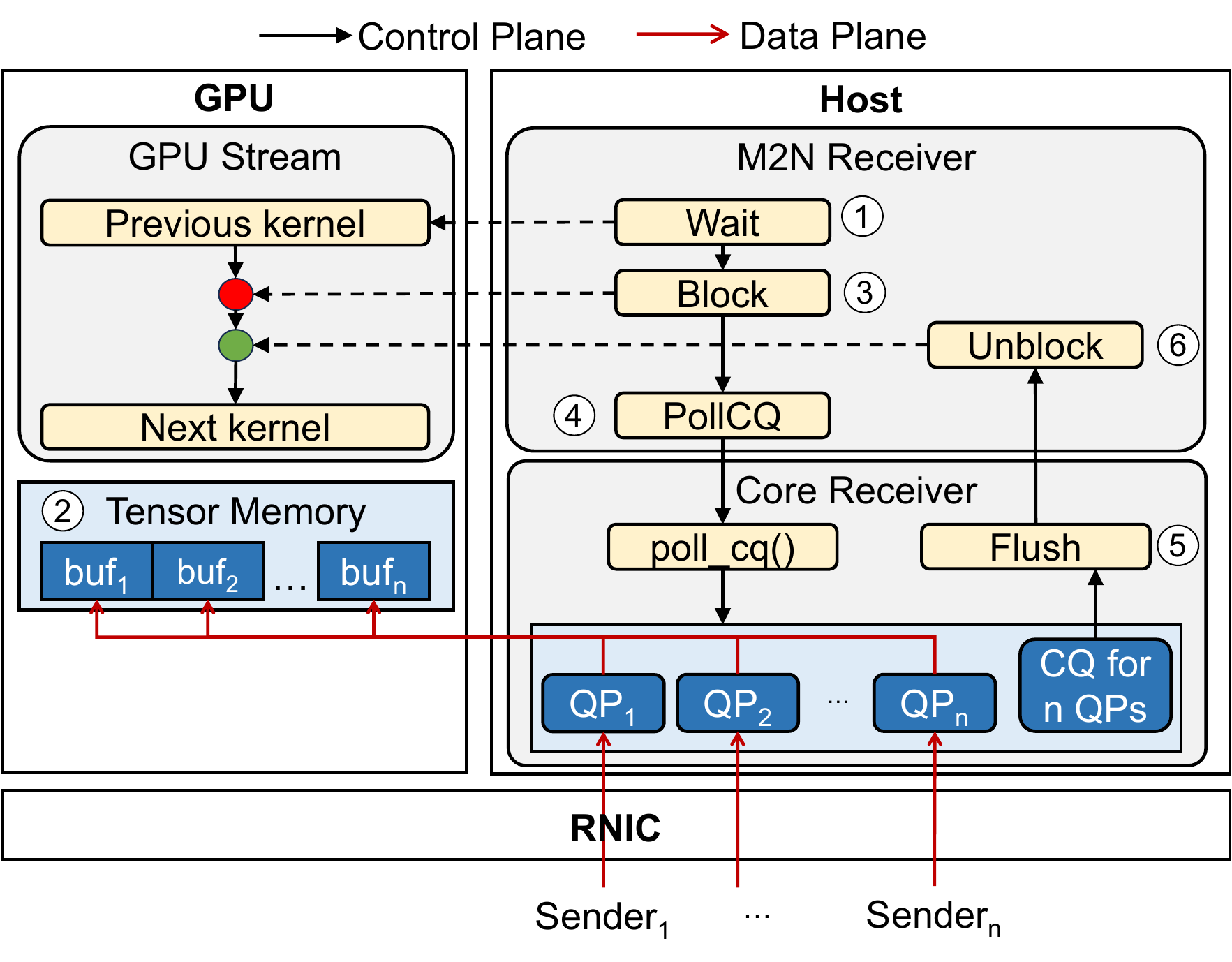}
    \vspace{-0.1in}
    \caption{M2N Receiver components and interactions.}
    \vspace{-0.1in}
    \label{fig:m2n:receiver}
\end{figure}

Based on these insights, we build our high-performance communication library with the goal of eliminating unnecessary GPU-to-CPU copies, group initialization/handling overhead, and GPU synchronization/memory accesses. Figures~\ref{fig:m2n:sender} and~\ref{fig:m2n:receiver} illustrate the sender and receiver architectures and their interactions within our M2N library.

\parabf{M2N Sender.} Figure~\ref{fig:m2n:sender} depicts the components of an M2N sender. In order to comply with the stream-oriented programming model, \ding{172} M2N senders utilize CUDA events \cite{nvidia-cuda-event-query} to wait for previous kernels and make sure that the \ding{173} pre-registered tensor to be transmitted is properly populated. Then, to ensure the next kernel in the stream starts after the transmission is complete, M2N senders utilize CUDA driver operations \cite{nvidia-driver-stream-wait} to \ding{174} block the stream. Once the stream is blocked, the data transmission proceeds in two steps: First, \ding{175} our in-house CPU communication library (denoted as \textit{Core Sender} in the figure) transmits the tensor efficiently using RDMA write with immediate~\cite{nvidia-dma-ops}. Second, to guarantee proper utilization of the registered tensor, \ding{176} the sender polls completions from the corresponding completion queue ~\cite{nvidia-dma-concepts}, confirming that data has been written to the remote buffers. Finally, M2N senders \ding{177} unblock the stream by updating a shared memory flag, allowing other kernels to continue reusing the registered memory. 
This design eliminates complex GPU synchronization, GPU-to-CPU copies, and group initialization overhead, all of which contribute to significant latency issues, especially for relatively small tensor sizes, as demonstrated in Figures~\ref{fig:eval:m2n_datasize} and~\ref{fig:eval:m2n_mn}.

\parabf{M2N Receiver.} Figure~\ref{fig:m2n:receiver} illustrates the components of an M2N Receiver. Just like its peer component, M2N receivers also need to adhere to the stream-oriented programming model.
Specifically, they \ding{172} wait on CUDA events to ensure that the \ding{173} pre-registered tensor is no longer in use, and \ding{174} block the stream to guarantee that subsequent kernels do not proceed until the operation is complete.
Once the stream is blocked, the data collection proceeds in two steps: First, receivers must verify that data has been successfully transmitted from the corresponding senders, which is efficiently achieved by \ding{175} polling the relevant completion queue. Second, to ensure data consistency at the GPU level, our in-house CPU communication library (denoted as \textit{Core Receiver} in the figure) leverages GDRCopy~\cite{nvidia-gdr-copy} and performs a \ding{176} flush operation~\cite{nccl-gdr-copy}. Finally, M2N receivers \ding{177} unblock the stream by updating a shared memory flag, allowing other kernels to continue utilizing the registered memory.
This simpler design eliminates the need for GPU-to-GPU copies and effectively reduces the GPU utilization overhead of receivers.

\parabf{Traffic-oriented optimizations.} We also introduce several traffic-oriented optimizations derived from our empirical observations during the scale-testing of our design.
\begin{itemize}[leftmargin=*]
    \item \textbf{High-priority ACKs.} We initially observed latency degradation in bidirectional communication with ping-pong pipeline parallelism. A detailed analysis revealed that ACK packets were often queued or transmitted with low priority (e.g., round-robin scheduling), leading to a large number of QP packets. Therefore, the receiving side experienced delays in responding to ACK packets, causing bottlenecks for M2N senders. To address this issue, we assign ACK packets to high-priority queues, isolating them from data packets, and fine-tuning the associated weight configurations empirically.
    
    \item \textbf{Congestion control fine-tuning.} We observed substantial latency degradations in unbalanced communication scenarios, where the amount of data to be sent varies significantly per receiver. To address this, we fine-tune our congestion control algorithms to minimize rate-limiting effects and allow faster convergence.
\end{itemize}

\parabf{Comparison with DeepEP.}
DeepEP~\cite{liu2024deepseek,deepseek-deepep,zhao2025insights} proposed by DeepSeek also optimizes network communication for large-scale expert parallel serving in MoE.
The key difference between our approach and DeepEP lies in the communication strategy: we leverage CPUs for inter-node communication, whereas DeepEP employs direct GPU-to-GPU communication without the involvement of CPU proxies~\cite{nvidia-ibgda}.
GPU-to-GPU communication approaches consume GPU compute resources on both the sender and receiver sides~\cite{nvidia-sm}.
In contrast, our CPU-to-CPU communication avoids the need for GPU compute resource allocation, thereby allowing compute kernels to fully utilize the GPU.
Moreover, GPU-to-GPU communication demands careful orchestration to mitigate contention between communication and computation kernels.
DeepEP employs custom PTX (assembly-like) instructions~\cite{nvidia-ptx} to minimize L2 cache usage---a resource shared between kernel types.
Our design, by comparison, does not require such low-level optimizations.
When handling requests within a single QP, the CPU achieves lower latency in our approach because its higher clock speed allows it to issue doorbells faster than the GPU.
However, the GPU offers stronger parallel processing capabilities, with multiple SMs able to independently manage QPs.
Consequently, DeepEP's approach can achieve a higher packet transmission rate at the cost of GPU SM resources and may yield better throughput when the packet size is very small.
In our scenario, the amount of data transferred between each sender-receiver pair typically reaches several hundred kilobytes (\S\ref{sec:evaluation:m2n}).
At this scale, a single-threaded CPU is sufficient to saturate the bandwidth.
If the number of experts increases further and the per-connection communication volume becomes smaller, leveraging the GPU’s superior parallel processing capabilities may offer greater advantages in terms of throughput.

\section{Implementation}
\label{sec:implementation}

\paraf{Fused kernels.}
To further improve efficiency and reduce latency, we implemented two types of fused kernels.
The first one is to overlap the communication of TP with the adjacent computation.
Although intra-node TP typically uses high-speed interconnects like NVLINK for communication, it still introduces non-negligible overhead.
To address this issue, we utilize Flux~\cite{chang2024flux} to fuse communication with the adjacent GEMM operation, such as implementing an all-gather and the following GEMM in a single kernel.
The second one is to fuse sequential memory-intensive operators.
MoE includes several sequences of small memory-intensive operations.
For example, attention nodes need to select top-k experts for each token after gating, compute intermediate results such as the number of tokens sent to each expert node and normalized token weights, and then perform data movement to scatter tokens to respective experts.
We optimize this process by fusing these steps with the gating computation, reducing both kernel launch and memory access.

\parabf{High-performance M2N communication library.} We built our communication library as a Pytorch extension \cite{torch-extension} in around 4900 and 5000 lines of C/C++ and Python code, respectively. Our library is supported by technologies such as GPUDirect~\cite{nvidia-gpudirect} and GDRCopy. 
We also carefully design network monitoring tools to delve into network and traffic-related optimizations.

\parabf{Load balance.}
In real-world traffic, the load across different experts can vary significantly.
To achieve load balancing between hot and cold experts, we deploy it with on-device redundancy based on expert popularity.
Specifically, we address the optimization problem of distributing $M$ experts across $N$ nodes in expert deployments. The objective is to minimize $\max_{j=1..N} C_j$, where $C_j = \sum_{i=1..M} x_{i,j} \cdot \max(a_i, K)$ represents the computational cost that corresponds to latency. $x_{i,j}$ denotes the allocation fraction, with $\sum_{j=1..N} x_{i,j} = 1$. $a_i$ represents the cost to calculate the active tokens of the expert $i$, and $K$ represents the lowest cost for the cold experts. The algorithm employs a greedy approximation strategy to solve this optimization problem and generate an expert plan, based on traffic within a previous time period.

\section{Evaluation}
\label{sec:evaluation}

\begin{table}[t]
    \centering
    \resizebox{1\linewidth}{!} {
    \begin{tabular}{ccccccc}
        \toprule
        \textbf{Model} & \textbf{\#Layers} & \textbf{Hidden Size} & \textbf{\#Experts} & \textbf{\topk} & \textbf{Intermediate Size} \\
        \midrule
        Mixtral-8$\times$22B & 56 & 6144 & 8 & 2 & 16384 \\
        DBRX & 40 & 6144 & 16 & 4 & 10752 \\
        Scaled-MoE & 48 & 8192 & 32 & 4 & 8192 \\
        \bottomrule
    \end{tabular}
    }
    \caption{Model configurations.}
    \vspace{-0.15in}
    \label{tab:evaluation:models}
\end{table}

\begin{figure*}[t!]
    \centering
    \includegraphics[width=0.9\linewidth]{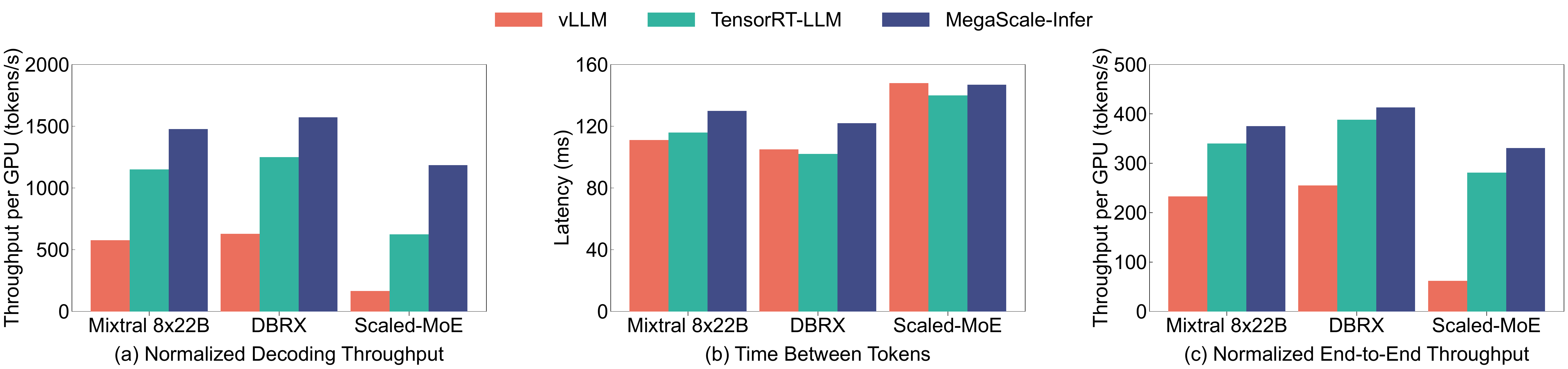}
    \vspace{-0.1in}
    \caption{Performance of \sysname on NVIDIA Ampere GPUs.}
    \vspace{-0.15in}
    \label{fig:eval:e2e_a800}
\end{figure*}

In this section, we first evaluate the end-to-end performance of \sysname against state-of-the-art LLM serving systems across various models and hardware configurations, including a heterogeneous environment.
Next, we demonstrate the effectiveness of high-performance M2N communication through micro-benchmarks.
Additionally, we conduct an ablation study to analyze the impact of disaggregated architecture and M2N optimization on \sysname's performance.
We also present the effectiveness of ping-pong pipeline parallelism and deployment plans, highlighting the benefits of deployment plan optimization.

\subsection{Experimental Setup}

\parabf{Testbed.}
We deploy \sysname across two distinct clusters.
The first cluster consists of eight nodes, each equipped with eight NVIDIA 80GB Ampere GPUs, 128 CPUs, 2 TB of host memory, and eight 200 Gbps Infiniband NICs.
GPUs within the same node are interconnected via 400GB/s NVLINK.
The second cluster is a heterogeneous setup, comprising two types of GPUs: NVIDIA H20 and L40S.
H20 nodes are equipped with 900GB/s NVLINK and four 400 Gbps NICs, while L40S nodes utilize PCIe for intra-node communication and two 400 Gbps NICs for inter-node communication.
As shown in Table~\ref{tab:design:hetero}, H20 offers higher memory capacity and bandwidth but has less computational power than L40S.

\parabf{Models and workload.}
We evaluate \sysname using Mixtral 8x22B~\cite{mixtral822}, DBRX~\cite{dbrx}, and a \scalemoe, which shares a similar structure but includes more experts.
They contain 141B, 132B, and 317B parameters, respectively.
The model configurations are detailed in Table~\ref{tab:evaluation:models}.
For all experiments, the data types for weights, activations, and KV cache are bfloat16.
We obtain a dataset from our production and use it as the experimental workload.
The median input and output length are 571 and 159 tokens, respectively.

\parabf{Baselines.}
We compare \sysname with two state-of-the-art serving systems: vLLM~\cite{kwon2023efficient} and TensorRT-LLM~\cite{trtllm}.
Both systems support popular techniques for LLM serving, including FlashAttention~\cite{dao2022flashattention}, PagedAttention~\cite{kwon2023efficient}, and continuous batching~\cite{yu2022orca}.
They primarily rely on tensor parallelism for distributed LLM serving, with TensorRT-LLM additionally supporting expert parallelism for expert layers.
For larger models requiring inter-node communication, both systems also support pipeline parallelism.
Due to GPU memory limits and large model sizes, serving Mixtral 8x22B and DBRX with these baselines requires a minimum of 8 GPUs, while \scalemoe necessitates multi-node deployment.
The Attention-FFN disaggregation architecture within \sysname naturally adapts to prefill/decoding (P/D) disaggregation, effectively preventing interference between these two phases. Existing baselines are still in the process of supporting or optimizing P/D disaggregation. To ensure a fair comparison and to accurately quantify the performance gains under P/D disaggregation, we evaluate all baselines and \sysname by temporally separating their prefill and decoding phases. Specifically, when measuring decoding throughput and time between tokens, we exclude the prefill phase and consider only the iteration time and output tokens of the decoding phase.

\parabf{Metrics.}
The primary objective of our work is to improve the efficiency of MoE inference during the decoding phase, which suffers from low GPU utilization due to its memory-bandwidth-bound nature. Enhancing decoding efficiency in MoE inference is the key to reducing overall serving costs. Therefore, we focus on maximizing the cost-normalized decoding throughput, subject to a specified time-between-tokens (TBT) latency constraint.
Specifically, for homogeneous deployment, we use per-GPU decoding throughput, i.e., tokens generated per second, excluding the first output token, divided by the number of GPUs, as the primary metric.
For heterogeneous deployment, we report per-cost decoding throughput.
Following prior work~\cite{zhong2024distserve}, we set the TBT requirement to 150 milliseconds.
We also report the mean time between tokens corresponding to the throughput results.
While our primary contribution lies in improving the efficiency of the decoding phase, we also report end-to-end throughput results that include the generation of the first output token.
This offers a more comprehensive evaluation of \sysname's performance improvements across both the prefill and decoding phases.
Notably, we observe that heterogeneous deployment also yields benefits for the prefill phase.
We present the throughput per unit power under heterogeneous deployment.
Additionally, we present the median and tail latency and throughput of M2N communication.

\begin{figure*}[t!]
    \centering
    \includegraphics[width=0.9\linewidth]{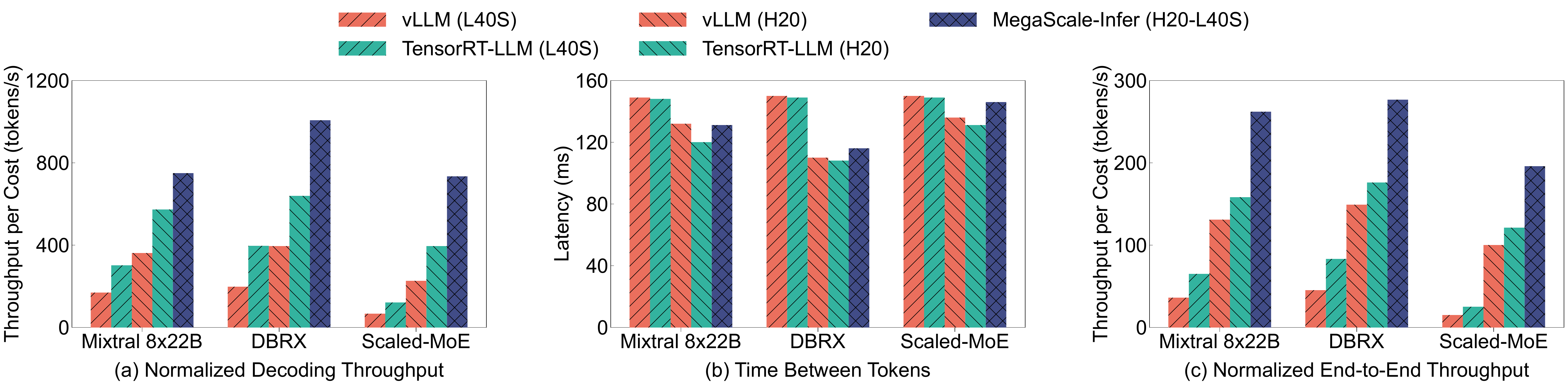}
    \vspace{-0.1in}
    \caption{Performance of \sysname on NVIDIA H20 and L40S GPUs.}
    \vspace{-0.1in}
    \label{fig:eval:e2e_hete}
\end{figure*}

\subsection{End-to-end Experiment}
\label{sec:evaluation:e2e}

\paraf{Homogeneous deployment.}
We first evaluate the performance of \sysname with different MoE models on NVIDIA 80GB Ampere GPUs.
Both vLLM and TensorRT-LLM serve Mixtral 8x22B and DBRX on a single node and serve \scalemoe across two nodes.
The results are presented in Figure~\ref{fig:eval:e2e_a800}.
Since vLLM deploys and serves the model as a whole, the batch size for the expert modules tends to be small, resulting in low GPU utilization.
TensorRT-LLM achieves higher throughput than vLLM through custom kernel optimizations, but it also adopts a holistic service approach, which means it cannot avoid the issue of low GPU utilization in the FFN modules.
By separating the attention and FFN modules, \sysname aggregates batched requests from multiple attention modules, which increases the FFN batch size.
This shift helps transition FFN from being memory-intensive to compute-intensive, thereby improving GPU utilization.
As a result, \sysname achieves $2.56\times$ and $1.28\times$ higher per-GPU decoding throughput than vLLM and TensorRT-LLM, as shown in Figure~\ref{fig:eval:e2e_a800}(a).
For \scalemoe, the expensive inter-node communication overhead, coupled with certain implementation limitations in a multi-node environment, results in even lower GPU utilization for the baselines.
In contrast, \sysname is deployed across multiple nodes for all models due to its disaggregated deployment, but it overlaps computation and communication through the design of ping-pong pipeline parallelism.
Consequently, \sysname improves the decoding throughput per GPU for \scalemoe by $7.11\times$ and $1.90\times$ compared to vLLM and TensorRT-LLM, respectively.

Figure~\ref{fig:eval:e2e_a800}(b) shows the mean time between tokens of \sysname and the baselines corresponding to the decoding throughput in Figure~\ref{fig:eval:e2e_a800}(a).
Due to the disaggregation of attention and expert modules, \sysname introduces cross-node communication at every layer, which affects latency. 
Ping-pong pipeline parallelism, while overlapping communication with computation across micro-batches, does not reduce the per-token latency for an individual micro-batch. Aiming for full GPU utilization, this approach may even incur additional latency, as indicated by constraint~\ref{formula:hide2}.
Nevertheless, \sysname effectively mitigates the communication overhead by employing a high-performance M2N communication library.
As a result, \sysname achieves a time between tokens comparable to those of the baseline systems.

Figure~\ref{fig:eval:e2e_a800}(c) shows the end-to-end per-GPU throughput including the prefill phase. As the prefill phase is predominantly compute-bound, our approach does not yield performance improvements for this stage under homogeneous deployments. Consequently, when the prefill phase is taken into account, the overall end-to-end performance gain is less pronounced compared to the decoding phase alone.
Nevertheless, \sysname still achieves up to a 1.18$\times$ improvement in throughput, demonstrating its effectiveness in end-to-end scenarios.

\parabf{Heterogeneous deployment.}
To demonstrate the benefits of \sysname under heterogeneous deployment, we build a cluster consisting of NVIDIA H20 and L40S GPUs and conduct experiments on it.
Since neither vLLM nor TensorRT-LLM supports heterogeneous deployment, we separately evaluate them on H20 and L40S.
To fully leverage the capacity of each GPU type, \sysname assigns H20 for attention modules and L40S for experts.
Figure~\ref{fig:eval:e2e_hete}(a) presents the performance measured by decoding throughput per unit cost.
Here we define cost with the normalized purchase price as shown in Table~\ref{tab:design:hetero}, which can easily be replaced by the rental price for cloud service users.
Due to the 48GB memory capacity of L40S, all models in both baselines require a multi-node setup.
In addition, the relatively weak intra-node and inter-node communication performance of the L40S leads to low GPU utilization for both vLLM and TensorRT-LLM.
In contrast, H20 is more suitable for LLM serving due to its large memory capacity, higher bandwidth, and faster communication.
As a result, vLLM and TensorRT-LLM achieve higher decoding throughput on H20.
However, the L40S also offers unique advantages, particularly its high compute power per unit cost, making it well-suited for executing compute-intensive tasks.
Our heterogeneous deployment simultaneously maximizes the advantages of the high bandwidth of H20 and the cost-effective compute power of L40S.
This results in an improvement of up to $3.24\times$ and $1.86\times$ on the unit cost decoding throughput compared to vLLM and TensorRT-LLM on H20, respectively.

\begin{figure}[t!]
    \centering
    \includegraphics[width=0.9\linewidth]{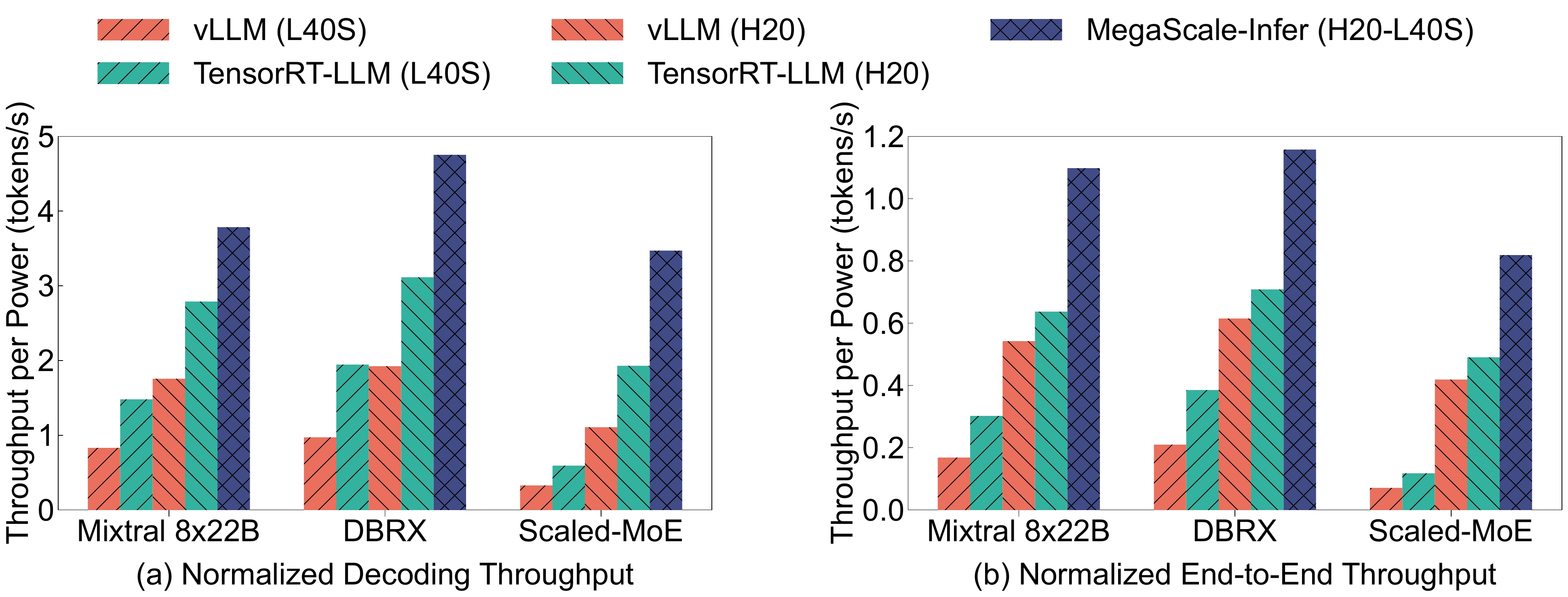}
    \vspace{-0.1in}
    \caption{Throughput per unit power of \sysname on NVIDIA H20 and L40S GPUs.}
    \vspace{-0.1in}
    \label{fig:eval:e2e_hete_power}
\end{figure}

\begin{figure*}[t!]
    \centering
    \includegraphics[width=0.9\linewidth]{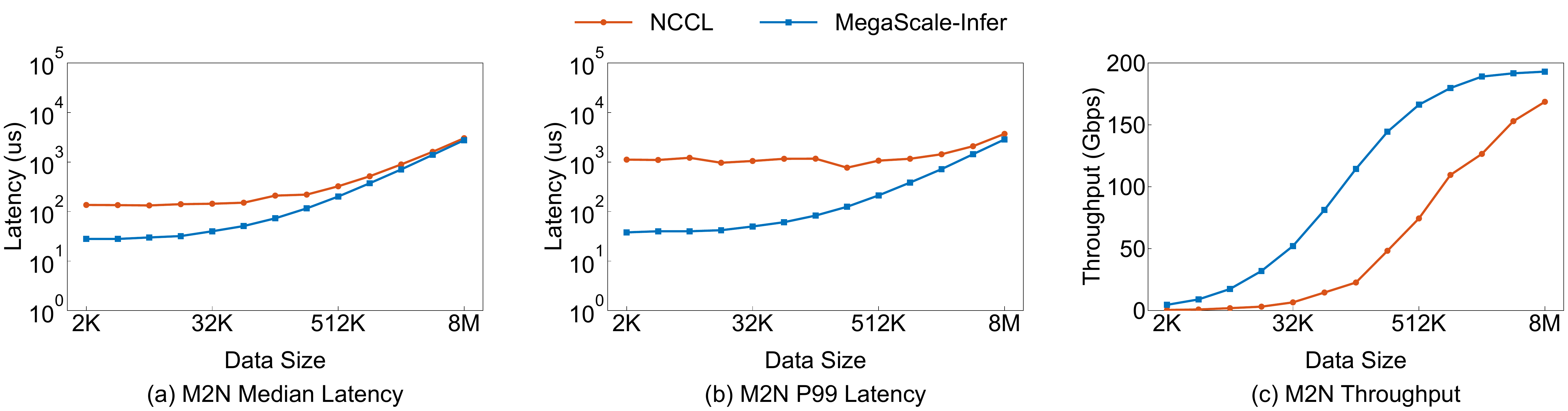}
    \vspace{-0.1in}
    \caption{Performance of M2N communication under different data sizes.}
    \vspace{-0.1in}
    \label{fig:eval:m2n_datasize}
\end{figure*}

\begin{figure*}[t!]
    \centering
    \includegraphics[width=0.9\linewidth]{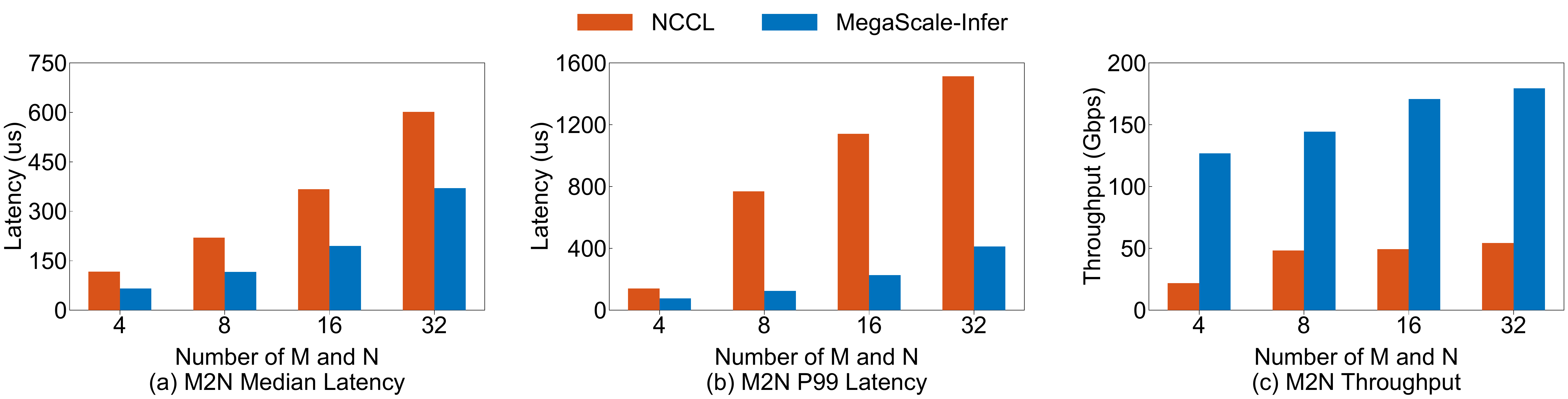}
    \vspace{-0.1in}
    \caption{Performance of M2N communication under different number of senders (M) and receivers (N).}
    \vspace{-0.1in}
    \label{fig:eval:m2n_mn}
\end{figure*}

The latency results corresponding to Figure~\ref{fig:eval:e2e_hete}(a) are shown in Figure~\ref{fig:eval:e2e_hete}(b).
Similar to the homogeneous deployment scenario, the mean time between tokens under heterogeneous deployment remains comparable to those of the baselines.
Moreover, when compared to the baselines deployed exclusively on L40S GPUs, our approach achieves slightly improved latency performance.

We further apply heterogeneous deployment to the prefill phase and report the end-to-end throughput results, including prefill computation, in Figure~\ref{fig:eval:e2e_hete}(c).
While heterogeneous deployment does not enhance resource utilization during the prefill phase, it effectively reduces inference costs by offloading expert computations to the more cost-efficient L40S GPUs.
As a result, when evaluating end-to-end performance across both the prefill and decoding phases, \sysname achieves up to a 1.66$\times$ improvement in throughput per unit cost compared to the baselines.
In summary, \sysname is particularly well-suited for heterogeneous deployment, as it significantly lowers inference costs across the entire serving pipeline.

We also evaluate the impact of heterogeneous deployment on power, and the results are presented in Figure~\ref{fig:eval:e2e_hete_power}.
Since the H20 and L40S GPUs each offer lower energy consumption per unit of bandwidth and compute, respectively, heterogeneous deployment across these two GPU types can also improve throughput per unit power.
This improvement is observed in both the decoding and prefill phases.
As shown in Figure~\ref{fig:eval:e2e_hete_power}(a) and (b), our system achieves 1.80$\times$ and 1.72$\times$ higher decoding and end-to-end throughput per unit power, respectively, compared to the baseline.

\subsection{Performance of M2N Communication}
\label{sec:evaluation:m2n}

We evaluate the performance of M2N communication under varying data sizes and different numbers of senders and receivers.
Each sender and receiver is a GPU equipped with a 200Gbps NIC.
The data size is defined as the bytes transmitted from one sender to one receiver.
In our MoE serving scenarios, data sizes range from hundreds of kilobytes.
For instance, serving Mixtral 8x22B with a micro-batch size of 128 and tensor parallelism of 2 for attention nodes requires each attention GPU to send an average of 196,608 bytes to each expert GPU, calculated as $\#tokens \times topk / \#experts \times hidden\_size \times sizeof(datatype)/TP = 128\times2/8\times6144\times2/2$.

Figure~\ref{fig:eval:m2n_datasize} illustrates how the latency and throughput of M2N vary with different data sizes.
In this experiment, we set the number of senders and receivers to 8.
Overall, \sysname achieves lower latency and higher throughput than NCCL across all data sizes.
NCCL incurs substantial overhead for small data sizes due to additional data copies and group operations.
To mitigate these overheads, we design and implement a highly optimized communication library, resulting in up to an 80.8\% reduction in median latency, as shown in Figure~\ref{fig:eval:m2n_datasize}(a).
Additionally, NCCL suffers from high tail latency due to its inherent instability.
We address this issue by eliminating GPU synchronization and group initialization.
As depicted in Figure~\ref{fig:eval:m2n_datasize}(b), we achieve up to 96.2\% reduction in P99 latency compared to NCCL.
These optimizations enable \sysname to improve throughput by up to $9.9\times$.
For commonly used data sizes in model serving, such as 256KB, \sysname achieves improvements of 68.2\% in median latency, 92.9\% in tail latency, and a $4.2\times$ increase in throughput.

We also scale the number of senders (M) and receivers (N) while keeping the data size fixed at 256 KB.
The results are shown in Figure~\ref{fig:eval:m2n_mn}.
\sysname consistently outperforms NCCL across all configurations.
As M and N increase, NCCL experiences greater instability, resulting in higher tail latency.
In contrast, our M2N library maintains stable performance through comprehensive traffic-oriented optimization, particularly congestion control fine-tuning.
This stability enables \sysname to reduce tail latency by 54.7\%-96.9\% and improve throughput by $3.3\times$-$5.8\times$.

\subsection{Ablation Study}

\begin{figure}[t]
    \centering
    \includegraphics[width=0.9\linewidth]{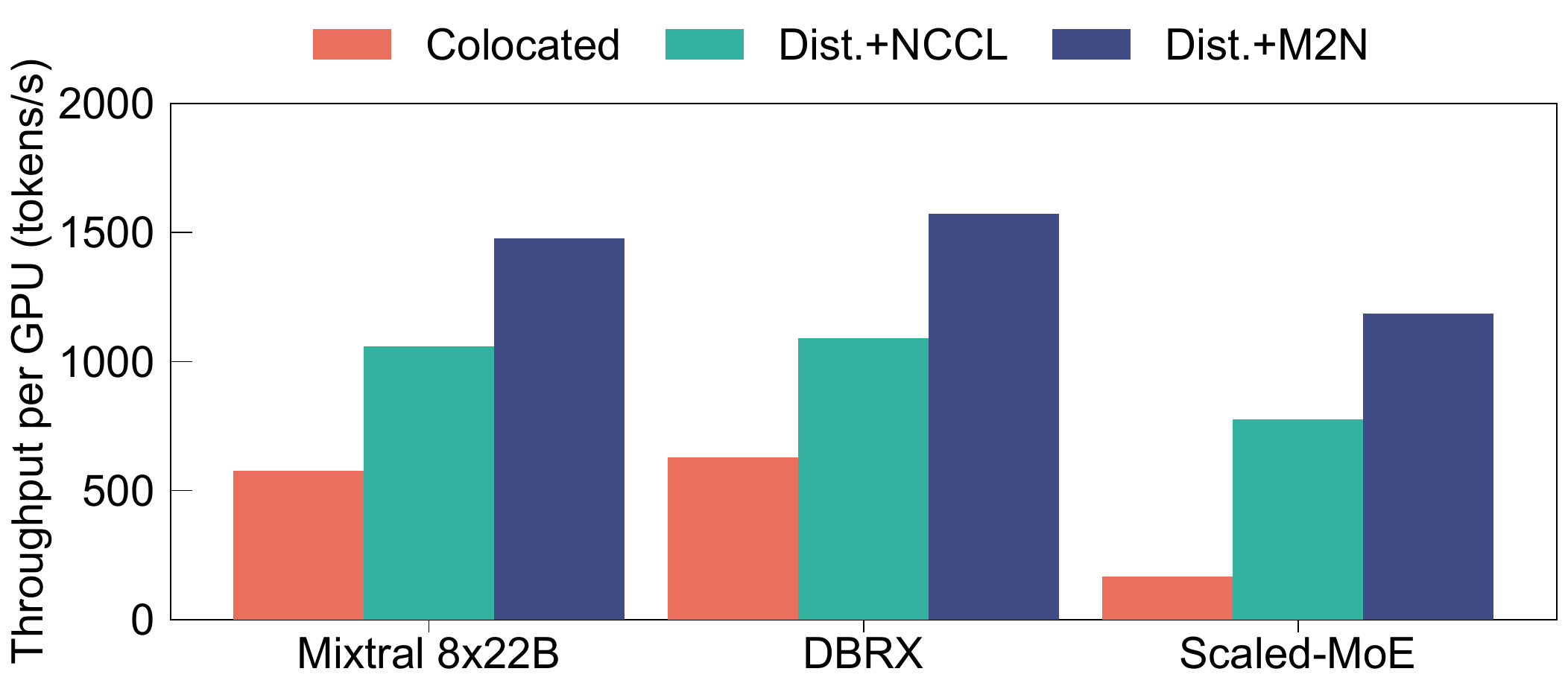}
    \vspace{-0.1in}
    \caption{Effectiveness of disaggregated expert parallelism and M2N optimization.}
    \vspace{-0.1in}
    \label{fig:eval:ablation}
\end{figure}

\paraf{Effectiveness of disaggregated expert parallelism and M2N optimization.}
Figure~\ref{fig:eval:ablation} illustrates the performance gains achieved on Ampere GPUs through disaggregated expert parallelism and M2N communication optimizations.
We select vLLM as the baseline, as it colocates attention and expert modules, and its kernel performance---particularly for matrix multiplication and attention operations---is comparable to our implementation.
By adopting a disaggregated architecture, requests from multiple attention modules can be aggregated, thereby increasing the effective batch size on the expert side and improving the computational efficiency of expert modules.
As a result, even when using NCCL as the backend for M2N communication, the disaggregated approach achieves up to a 4.66$\times$ throughput improvement over the colocated baseline.
By leveraging our optimized M2N communication library, we further reduce communication overhead, enabling the M2N communication time of a single micro-batch to fall below its computation time (satisfying constraint~\ref{formula:hide1}).
As a result, communication can be fully overlapped with computation through the use of the ping-pong pipeline, leading to an additional throughput improvement of up to 1.53$\times$.

\begin{figure}[t]
    \centering
    \includegraphics[width=0.9\linewidth]{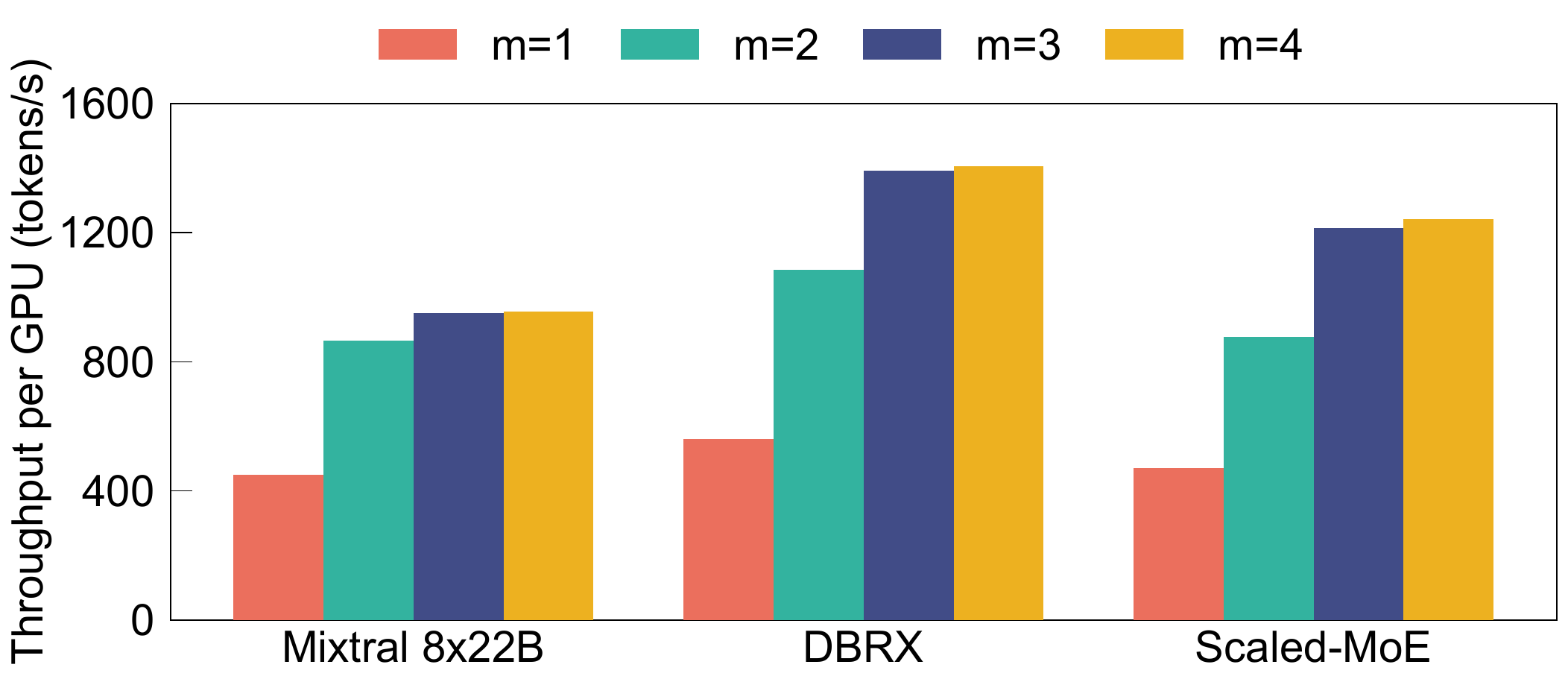}
    \vspace{-0.1in}
    \caption{Normalized decoding throughput under different numbers of micro-batch.}
    \vspace{-0.1in}
    \label{fig:eval:pingpong}
\end{figure}

\parabf{Effectiveness of ping-pong pipeline parallelism.}
First, we conduct an ablation study by varying the number of micro-batches ($m$) while keeping the micro-batch size constant.
To fully demonstrate the benefits of ping-pong pipeline parallelism, we adopt the optimal deployment plan where the computation times of attention and FFN modules are nearly balanced.
Figure~\ref{fig:eval:pingpong} presents the evaluation results on Ampere GPUs.
When $m=1$, ping-pong pipeline parallelism is disabled, leading to idle periods for either attention or FFN module while the other is computing.
This results in relatively low decoding throughput across all models.
Increasing $m$ from 1 to 2 enables both modules to simultaneously process two micro-batches in a ping-pong manner, significantly reducing idle time and improving throughput by 1.9$\times$.
While $m=2$ is ideally sufficient to achieve high GPU utilization, inter-node communication overhead remains a significant factor.
By increasing $m$ to 3, we enable the overlapping of communication and computation, resulting in throughput improvements of 1.10$\times$, 1.28$\times$, and 1.38$\times$ for Mixtral 8x22B, DBRX, and \scalemoe, respectively.
Larger models require more GPUs for serving, leading to increased communication overhead.
Consequently, increasing $m$ provides more significant benefits for larger models.
Given the high network bandwidth in our testbed, further increasing $m$ yields only marginal improvements.

\begin{figure}[t]
    \centering
    \subfloat[]{
        \includegraphics[width=0.45\linewidth]{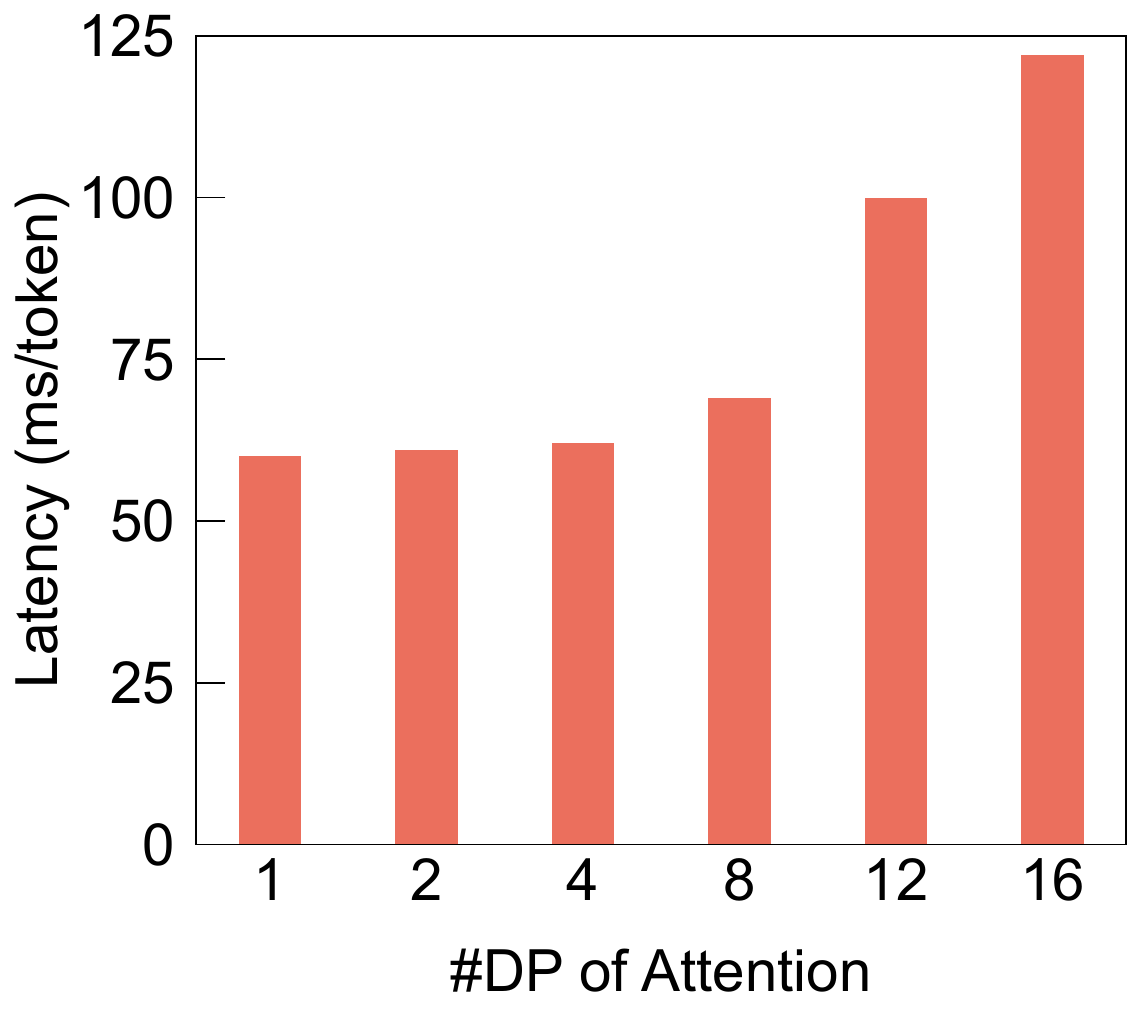}
        \label{fig:eval:dp_lat}
    }
    \subfloat[]{
        \includegraphics[width=0.45\linewidth]{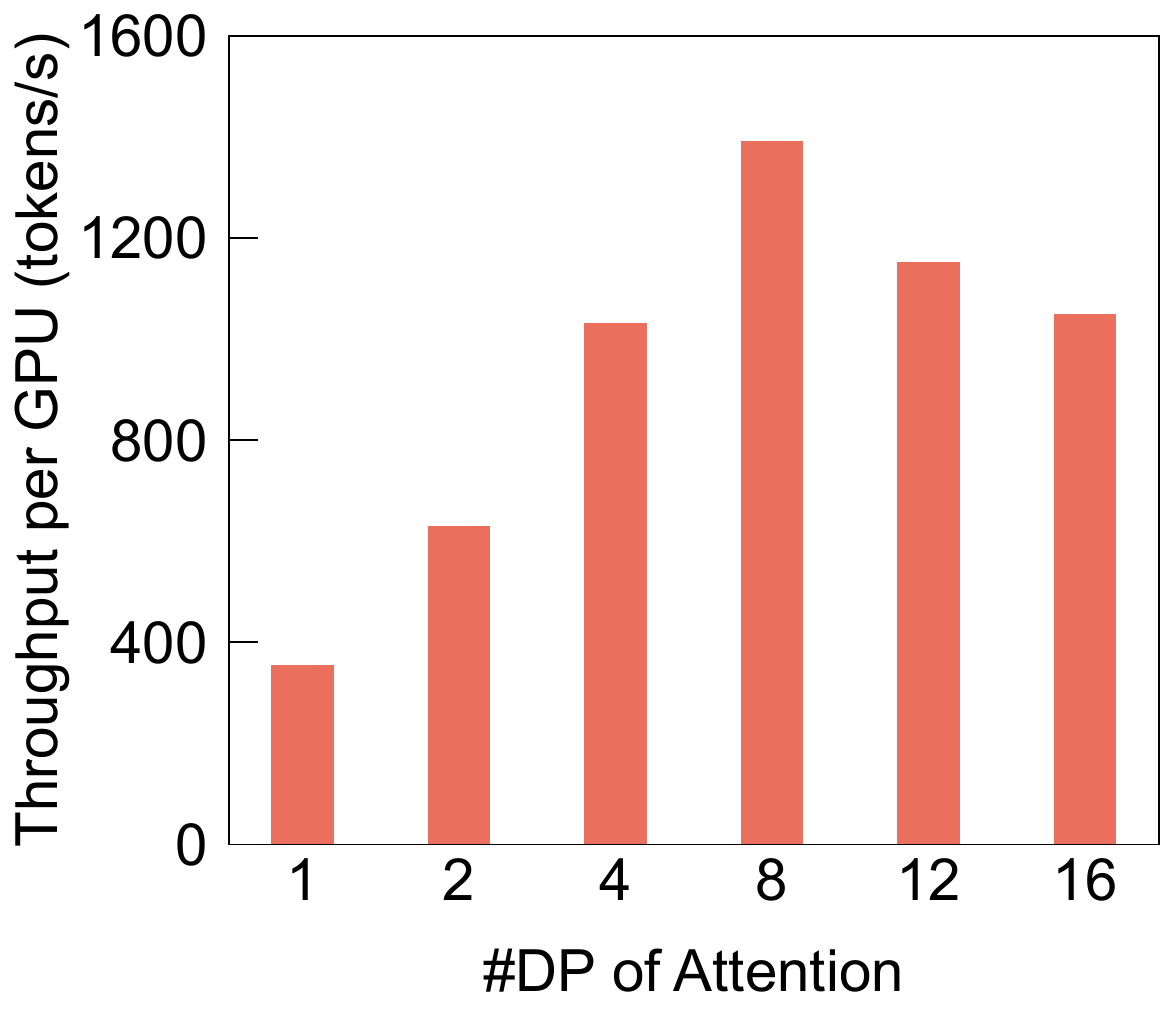}
        \label{fig:eval:dp_tput}
    }
    \vspace{-0.1in}
    \caption{Performance of DBRX under different DP degree of attention. (a) Per output token latency. (b) Decoding throughput normalized by number of GPUs.}
    \vspace{-0.1in}
    \label{fig:eval:dp}
\end{figure}

\parabf{Influence of deployment plan.}
We further investigate the impact of the deployment plan using DBRX as a case study by varying the degree of data parallelism (DP), i.e., the number of replicas for the attention nodes.
The number of micro-batches is fixed at 3 to maximize the benefits of ping-pong pipeline parallelism.
Figure~\ref{fig:eval:dp} shows the resulting latency and throughput.
With a small DP degree, each expert processes fewer tokens, leading to a shorter computation time for the FFN module compared to the attention.
As a result, expert nodes experience significant idle time, even with ping-pong pipeline parallelism employed.
As shown in Figure~\ref{fig:eval:dp_lat}, the latency remains constant as the DP degree increases from 1 to 4, suggesting that the attention module is the bottleneck.
Meanwhile, Figure~\ref{fig:eval:dp_tput} demonstrates linear throughput scaling within this range, further confirming that the bottleneck is in the attention module.
When the DP degree reaches 8, the computation times for both attention and FFN become roughly equal, allowing both modules to stay busy during inference.
As seen in Figure~\ref{fig:eval:dp}, the latency in this case is similar to that with a lower DP degree, while the normalized decoding throughput reaches its peak.
As the DP degree continues to increase, expert nodes are assigned more tokens, causing the bottleneck to shift from attention to experts.
This leads to higher latency and reduced normalized throughput, as attention nodes experience significant idle time.
This experiment showcases the importance of optimizing the deployment plan.
Only certain deployment plans can minimize idle time and maximize GPU utilization.

\section{Deployment Experience}
\label{sec:experience}

\begin{figure*}[t]
    \centering
    \subfloat[]{
        \includegraphics[width=0.323\linewidth]{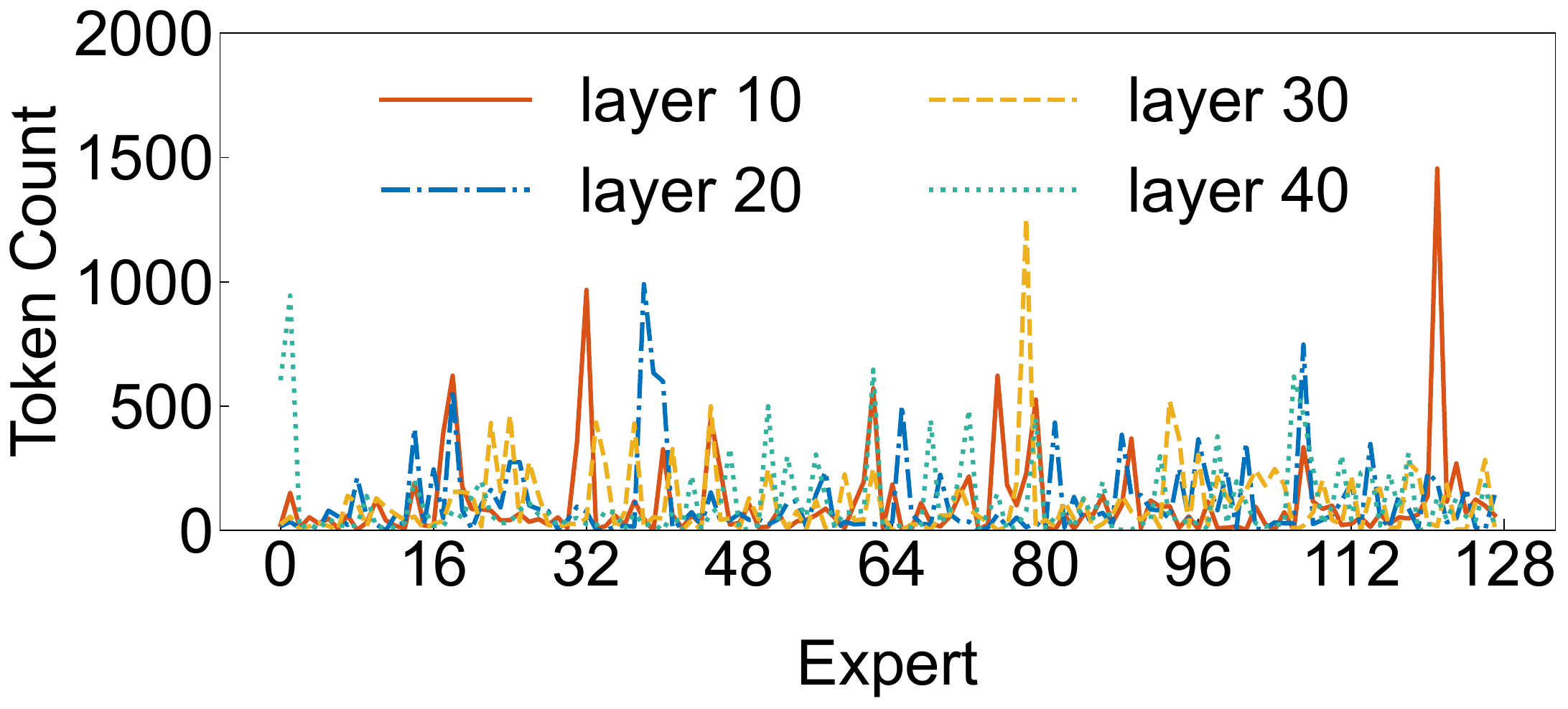}
        \label{fig:exp:ep}
    }
    \subfloat[]{
        \includegraphics[width=0.31\linewidth]{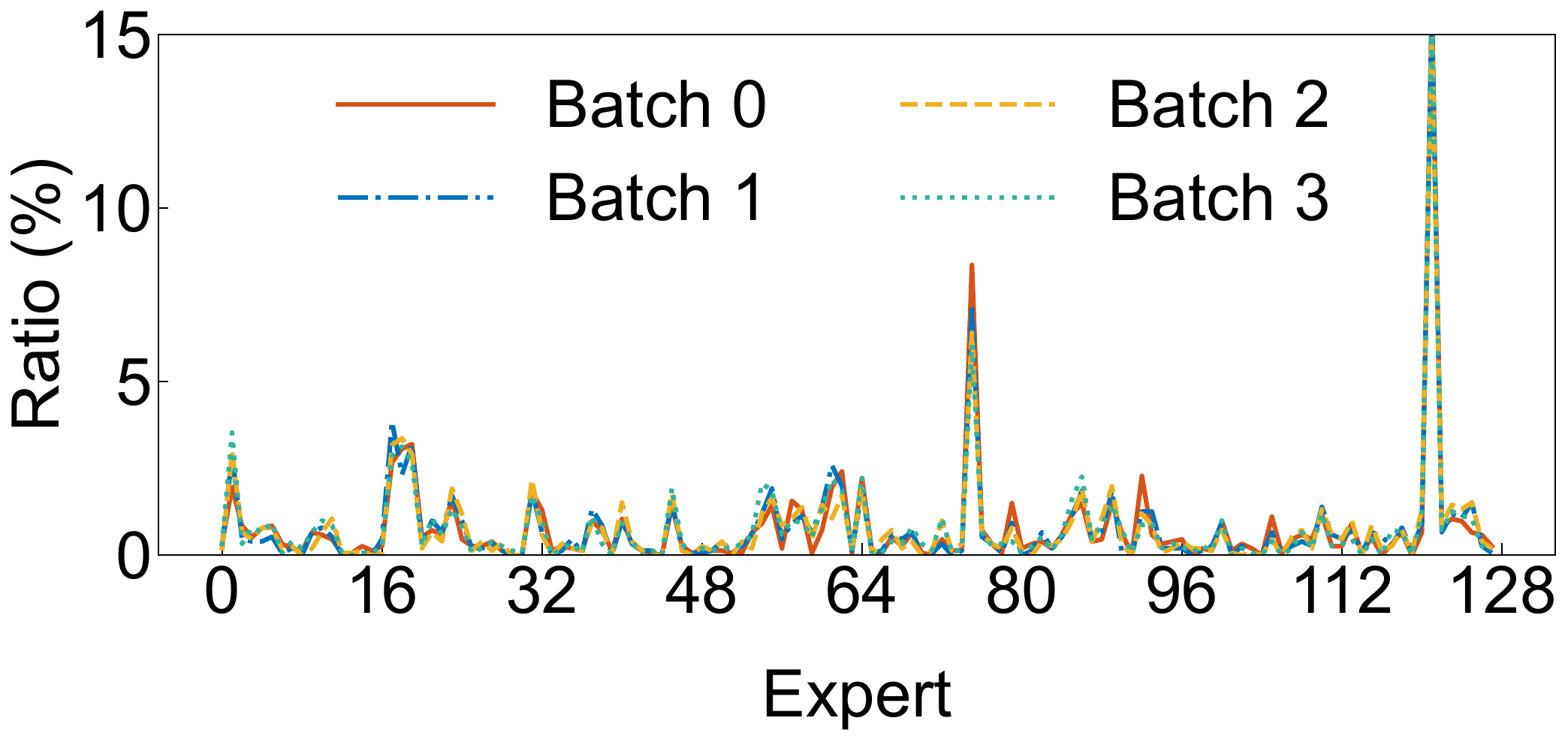}
        \label{fig:exp:dec}
    }
    \subfloat[]{
        \includegraphics[width=0.31\linewidth]{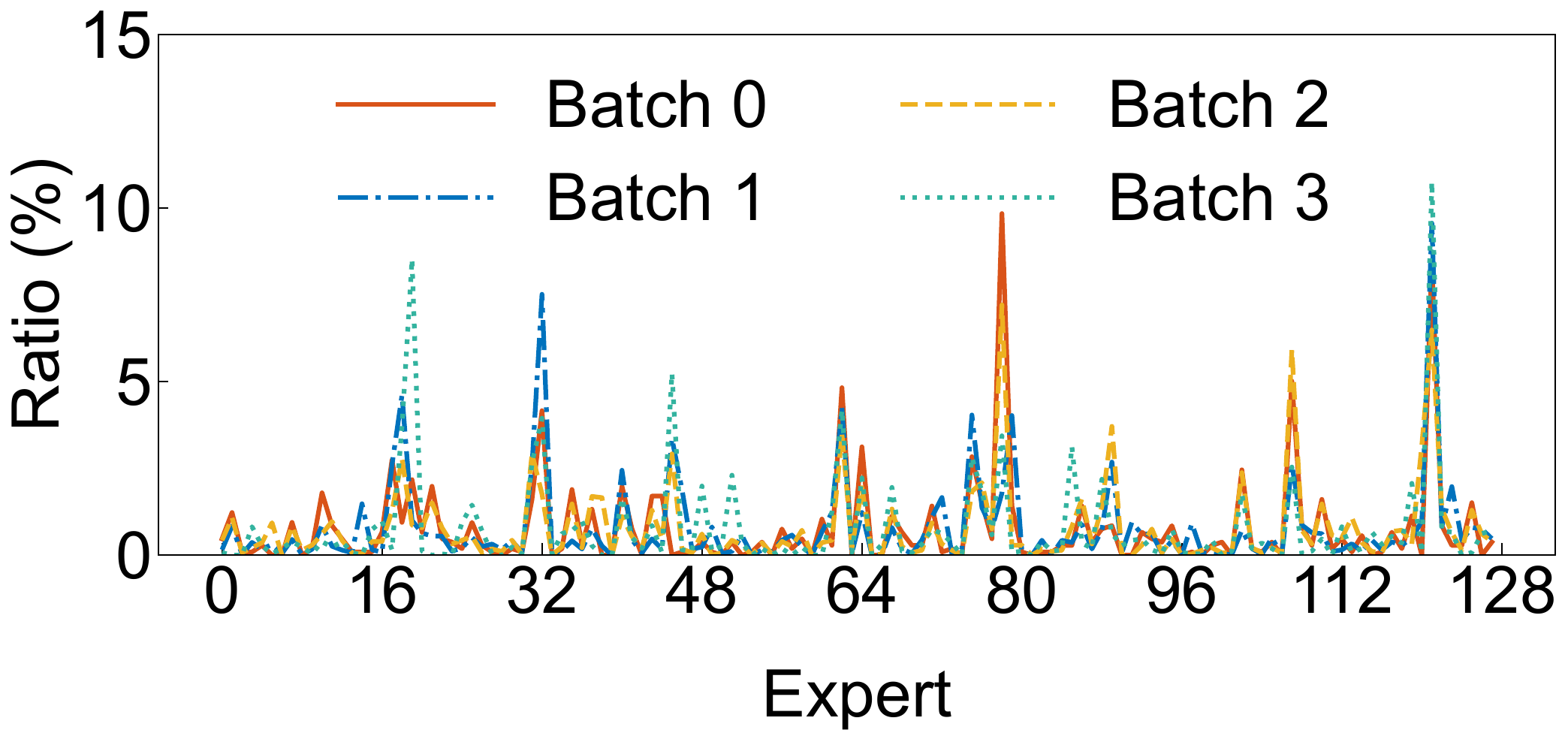}
        \label{fig:exp:ctx}
    }
    \vspace{-0.1in}
    \caption{Expert load distribution in real traffic. (a) Received token count of each expert in a batch. (b) Received token ratio of each expert during decoding. (c) Received token ratio of each expert during prefill.}
    \vspace{-0.1in}
    \label{fig:exp:expert}
\end{figure*}

\sysname has been deployed in the company's production inference services and is operating on a cluster with nearly 10,000 GPUs. Under heterogeneous deployment, it reduces the cost of serving the same traffic by 1.5--2.0$\times$, depending on the workload characteristics.

\parabf{Expert balance.} In real-world deployment, we gained additional insights into expert load distribution.
Figure~\ref{fig:exp:ep} illustrates the number of tokens processed by each expert across four model layers for a single batch.
There is a significant imbalance in the load distribution, with some experts being substantially hotter than others.
Further analysis of different phases reveals additional patterns.
As shown in Figures~\ref{fig:exp:dec} and~\ref{fig:exp:ctx}, we sample four batches executed within a short time window and measure the load on all experts in a specific layer.
During the decoding phase, expert load remains relatively stable across batches, whereas in the prefill phase, it fluctuates more significantly.
These observations motivate us to adopt a static or periodic expert load balancing strategy during decoding, while employing more frequent expert plan adjustments in the prefill phase.

\parabf{Attention balance.} We also observe load imbalance on the attention side.
Due to variations in sequence lengths, attention nodes can experience significantly different computation times even when processing batches of the same size.
Since \sysname performs frequent synchronization across all nodes, such imbalances can introduce bubbles and degrade overall system efficiency.
To mitigate this issue, we profile the runtime of key operators under varying sequence lengths and batch sizes to estimate the computation time of a given batch.
We then compose the batch on each attention node to match a predefined target execution time, thereby balancing the workload across nodes.

\section{Related Work}
\label{sec:related}

\paraf{LLM serving.}
Recently, numerous works have been proposed to optimize LLM inference.
Orca~\cite{yu2022orca} introduces iteration-level scheduling to improve throughput.
vLLM~\cite{kwon2023efficient} enhances KV cache management through PagedAttention for greater efficiency.
Sarathi-Serve~\cite{agrawal2024taming} addresses the throughput-latency tradeoff by splitting prefill requests into chunks and batching them with ongoing decoding requests.
LoongServe~\cite{wu2024loongserve} leverages elastic sequence parallelism to efficiently serve long-context requests in dynamic workloads.
For serving multiple instances, Llumnix~\cite{sun2024llumnix}, ServerlessLLM~\cite{fu2024serverlessllm}, and dLoRA~\cite{wu2024dlora} propose request migration techniques to enable load balancing and reduce latency.
However, their approaches primarily focus on dense models, often overlooking the distinct challenges introduced by the sparsity of large-scale MoE models.

\parabf{Resource disaggregation.} 
Disaggregating hardware resources into separate resource pools allows for independent scaling, resulting in more efficient deployments.
Several systems~\cite{guo2023mira, shan2018legoos} adopt this approach.
In the context of LLM serving, the distinct characteristics of the prefill and decoding phases make their disaggregation a widely used solution~\cite{zhong2024distserve, patel2023splitwise, hu2024inference, strati2024dejavu, qin2024mooncake}.
\sysname also employs this approach and further optimizes MoE serving efficiency during decoding by disaggregating attention and FFN modules.
FASTDECODE~\cite{he2024fastdecode}, Lamina~\cite{chen2024efficient}, and MoE-Lightning~\cite{cao2024moe} offload attention computation to cheaper devices, such as CPU, during decoding.
However, offloading results in higher latency, and the challenges posed by MoE's sparsity remain unresolved.

\parabf{MoE optimization.}
MoE has gained popularity for its ability to reduce computational complexity~\cite{lepikhin2020gshard, xu2021gspmd, rajbhandari2022deepspeed, fedus2022switch, liu2024deepseek}.
Currently, two primary considerations in optimizing MoE training and inference are load balancing~\cite{cui2023optimizing, li2023accelerating} and efficient communication~\cite{hwang2023tutel, li2023accelerating, rajbhandari2022deepspeed, liu2023janus}.
Serving large-scale MoE with limited resources also demands efficient offloading and preloading~\cite{hwang2024pre, cao2024moe}.
In this work, we focus on large-scale distributed serving, addressing the unique inefficiencies introduced by MoE's sparsity through disaggregation.

\parabf{Collective communication for ML.}
Distributed machine learning jobs heavily rely on high-performance collective communications, such as all-reduce and all-to-all, to achieve high throughput and low latency.
NVIDIA NCCL~\cite{nccl} is the most popular collective communication library in both industry and academia.
SCCL~\cite{cai2021synthesizing}, TACCL~\cite{shah2023taccl}, and TE-CCL~\cite{liu2024rethinking} propose automatic synthesis of optimal collective communication algorithms tailored to distinct hardware and topologies.
CoCoNET~\cite{jangda2022breaking} and Centauri~\cite{chen2024centauri} improve performance by overlapping communication with computation in distributed machine learning.
The disaggregation of attention and FFN in MoE necessitates a new form of collective communication, i.e., M2N.
We identify and eliminate the overhead and instability present in existing solutions.
\section{Conclusion}
\label{sec:conclusion}

In this paper, we present \sysname, a system that disaggregates the attention and FFN modules to enhance the efficiency and cost-effectiveness of large-scale MoE serving.
Leveraging this disaggregation architecture, \sysname builds the optimal deployment plan with a ping-pong parallelism strategy and a high-performance M2N communication library.
The evaluation results across diverse models and hardware demonstrate that \sysname achieves up to 1.9$\times$ throughput improvement over state-of-the-art systems, highlighting the effectiveness of our design and implementation.

\clearpage

\bibliographystyle{plainnat}
\bibliography{reference}




\end{document}